\documentclass[sigconf]{acmart}
\usepackage{xcolor}
\usepackage{multirow}
\usepackage{booktabs} 
\usepackage{amsmath}
\usepackage{graphicx}
\usepackage{caption}
\usepackage{graphics}
\usepackage{rotating}
\usepackage{tabularx}
\usepackage{wrapfig}
\usepackage{listings}
\usepackage{color, colortbl}
\usepackage{balance} 
\usepackage{lscape}
\usepackage{hyperref}
\usepackage{xspace}
\usepackage{framed}
\usepackage{syntax}
\usepackage{soul}
\usepackage{flushend}
\usepackage{caption}
\usepackage{subcaption}
\usepackage{tabularx,ragged2e}
\usepackage{booktabs}
\usepackage{amsmath}
\usepackage{algorithm}
\usepackage[noend]{algpseudocode}
\usepackage{url}

\usepackage{tabularx}
\usepackage{array}
\usepackage{colortbl}
\usepackage{varwidth}
\usepackage[justification=centering]{caption}
\usepackage[tikz]{bclogo}
\usepackage[braket]{qcircuit}

\newcounter{ec}

\newtheorem{example}[ec]{Example}

\makeatletter
\def\BState{\State\hskip-\ALG@thistlm}
\makeatother
\definecolor{sandstorm}{rgb}{0.93, 0.84, 0.25}
\definecolor{codegreen}{rgb}{0,0.6,0}
\definecolor{codegray}{rgb}{0.5,0.5,0.5}
\definecolor{codepurple}{rgb}{0.58,0,0.82}
\definecolor{backcolour}{rgb}{0.95,0.95,0.92}
\definecolor{Gray}{gray}{0.1}
\definecolor{Blanched Almond}{rgb}{1.0, 0.92, 0.8}
\definecolor{darksalmon}{rgb}{0.91, 0.59, 0.48}
\definecolor{AliceBlue}{rgb}{0.94, 0.97, 1.0}
\definecolor{darkred}{rgb}{0.55, 0.0, 0.0}
\definecolor{Steelblue}{rgb}{0.27, 0.51, 0.71}

\newcolumntype{Y}{>{\raggedleft\arraybackslash}X}

\lstdefinestyle{mystyle}{
	backgroundcolor=\color{AliceBlue},   
	commentstyle=\color{codegreen},
	keywordstyle=\color{magenta},
	numberstyle=\tiny\color{codegray},
	stringstyle=\color{codepurple},
	basicstyle=\footnotesize,
	breakatwhitespace=false,         
	breaklines=true,                 
	captionpos=b,                    
	keepspaces=true,                 
	numbers=left,                    
	numbersep=5pt,                  
	showspaces=false,                
	showstringspaces=false,
	showtabs=false,                  
	tabsize=2,
	columns=fullflexible,
	moredelim=[is][\color{blue}\bfseries\underbar]{@}{@},
	escapechar=\@,
	mathescape=true
}

\newcommand{\ak}{\textit{Auto-Keras}\xspace}
\newcommand{\om}{\textit{Original Manas}\xspace}
\newcommand{\tm}{\textit{Transformed Manas}\xspace}

\newcommand{\gh}{\textit{GitHub}\xspace}
\newcommand{\sof}{\textit{StackOverflow}\xspace}
\newcommand{\kg}{\textit{Kaggle}\xspace}
\newcommand{\keras}{\textit{Keras}\xspace}
\newcommand{\manas}{\textit{Manas}\xspace}

\newcommand{\figref}[1]{Figure~\ref{#1}}

\newcommand{\etal}{{et al.}\xspace}

\newcommand{\nodata}{8\xspace}

\newcommand{\errorrate}{{\bf 17.6\%}\xspace} 
\newcommand{\mse}{{\bf 0.0\%}\xspace}
 
\newcommand{\depth}{{\bf 75.7\%}\xspace} 
\newcommand{\width}{{\bf 93.0\%}\xspace} 
\newcommand{\speed}{{\bf 247.1\%}\xspace} 
\newcommand{\depthreg}{{\bf 74.4\%}\xspace} 
\newcommand{\widthreg}{{\bf 89.1\%}\xspace} 
\newcommand{\speedreg}{ {\bf 66.3\%}\xspace}

\newcounter{NumObservations}
\stepcounter{NumObservations}
\newcommand*\circled[1]{\tikz[baseline=(char.base)]{
		\node[shape=circle,fill,inner sep=1pt] (char) {\textcolor{white}{#1}};}}

\AtBeginDocument{%
  \providecommand\BibTeX{{%
    \normalfont B\kern-0.5em{\scshape i\kern-0.25em b}\kern-0.8em\TeX}}}

\copyrightyear{2022}
\acmYear{2022}
\setcopyright{rightsretained}
\acmConference[ICSE '22]{44th International Conference on Software Engineering}{May 21--29, 2022}{Pittsburgh, PA, USA}
\acmBooktitle{44th International Conference on Software Engineering (ICSE '22), May 21--29, 2022, Pittsburgh, PA, USA}
\acmDOI{10.1145/3510003.3510052}
\acmISBN{978-1-4503-9221-1/22/05}

\begin{document}

\title{Manas: Mining Software Repositories to Assist AutoML}

\author{Giang Nguyen}
\email{gnguyen@iastate.edu}
\affiliation{
  \institution{Dept. of Computer Science, Iowa State University}
  \city{Ames}
  \state{IA}
  \country{USA}
  \postcode{50011}
}

\author{Md Johirul Islam}
\email{mislam@iastate.edu}
\authornote{This work was done when Md Johirul Islam was at Iowa State University.}
\affiliation{
  \institution{Dept. of Computer Science, Iowa State University}
  \city{Ames}
  \state{IA}
  \country{USA}
  \postcode{50011}
}

\author{Rangeet Pan}
\email{rangeet@iastate.edu}
\affiliation{
  \institution{Dept. of Computer Science, Iowa State University}
  \city{Ames}
  \state{IA}
  \country{USA}
  \postcode{50011}
}

\author{Hridesh Rajan}
\email{hridesh@iastate.edu}
\affiliation{
  \institution{Dept. of Computer Science, Iowa State University}
  \city{Ames}
  \state{IA}
  \country{USA}
  \postcode{50011}
}

\renewcommand{\shortauthors}{Giang Nguyen, Md Johirul Islam, Rangeet Pan, and Hridesh Rajan}

\begin{abstract}
Today deep learning is widely used for building software.
A software engineering problem with deep learning is that finding an appropriate 
convolutional neural network (CNN) model for the task can be a challenge for developers. 
Recent work on AutoML, more precisely neural architecture search (NAS), embodied 
by tools like \ak aims to solve this problem by essentially viewing it as a search problem 
where the starting point is a default CNN model, and mutation of this 
CNN model allows exploration of the space of CNN models to 
find a CNN model that will work best for the problem. 
These works have had significant success in producing high-accuracy CNN models. 
There are two problems, however.  
First, NAS can be very costly, often taking several hours to complete. 
Second, CNN models produced by NAS can be very complex that makes it 
harder to understand them and costlier to train them. 
We propose a novel approach for NAS, where instead of starting from a default 
CNN model, the initial model is selected from a repository of models extracted 
from \gh. 
The intuition being that developers solving a similar problem may have developed 
a better starting point compared to the default model. 
We also analyze common layer patterns of CNN models in the wild to understand 
changes that the developers make to improve their models. 
Our approach uses commonly occurring changes as 
mutation operators in NAS.
We have extended \ak to implement our approach. 
Our evaluation using \nodata top voted problems from \kg for tasks including image classification and image regression shows that given the same search time, without loss of accuracy, \manas produces models with {\bf 42.9\%} to {\bf 99.6\%} fewer number of parameters than \ak' models. Benchmarked on GPU, \manas' models train {\bf 30.3\%} to {\bf 641.6\%} faster than \ak' models.

\end{abstract}



\maketitle




	\section{Introduction}
\label{sec:intro}
\begin{figure*}[t]
	\centering
	\begin{subfigure}{0.35\textwidth}
		\hspace{-8em}
		\includegraphics[keepaspectratio = True, scale= .33]{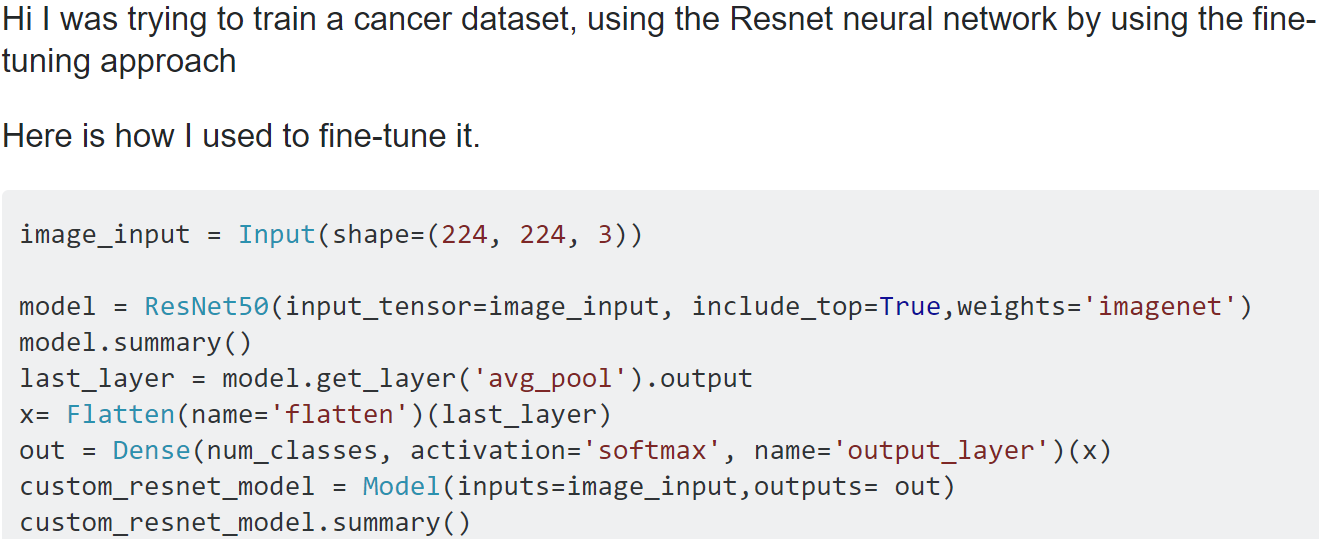}
		\caption{Problem}
		\label{fig:question}
	\end{subfigure}%
	\begin{subfigure}{0.33\textwidth}
		\hspace{1em}
		\includegraphics[keepaspectratio = True, scale= .29]{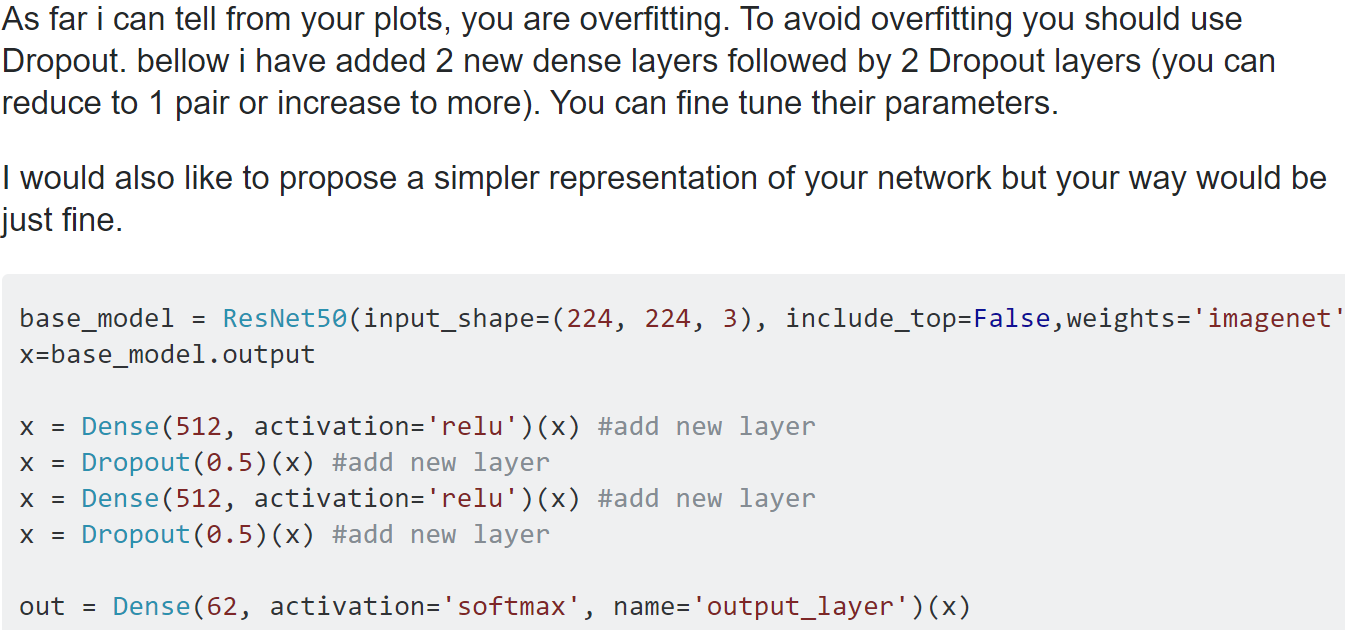}
		\caption{Solution}
		\label{fig:answer}
	\end{subfigure}
	\vspace{-0.8em}
	\caption{An example from \sof showing the necessity to change model architecture}
	\label{fig:motivation}
	\vspace{-0.6em}
\end{figure*}

An increasingly larger number of software systems today are including 
deep learning.
Deep learning uses a convolutional neural network (CNN) model, 
essentially a graph with weighted edges and nodes that are 
weighted functions, to convert inputs to the output.
As more software systems incorporate deep learning, more 
software developers have to design and train CNN models. 
Designing a CNN model is very difficult, and developers
often struggle leading to bugs. 
Model bugs are frequent bugs in CNN programs~\cite{islam19,islam20repairing}.

Recent work on neural architecture search (NAS) aims to solve this problem~\cite{DBLP:conf/iclr/ZophL17}. 
NAS techniques start from a collection of default CNN models and search for a suitable model
for the problem. 
The search space is defined by the collection of default models and a collection of mutation operators that are used to modify
CNN models to create new candidates. 
NAS techniques have been implemented in industrial-strength tools such as \ak~\cite{jin2019auto}. 
NAS techniques face two problems. 
First, NAS can be very costly, e.g., \ak takes 8-12 hours on high performance
machines to search for models with reasonable accuracy (90+\%).
The accuracy drops rapidly if the search time is reduced.
Second, CNN models produced by NAS can be very complex 
that makes it harder to understand them for maintenance, 
and costlier to train and retrain them.

We introduce \manas (Mining Assisted Neural Architecture 
Search) to alleviate 
the limitations of NAS. 
The fundamental intuition behind \manas is that mining and using the 
hand-developed models that are available in open-source repositories as 
default models or starting point of search can help NAS leverage human developer efforts. 
\manas applies this intuition in two ways. 
First, hand-developed models are mined to search for a better starting point for NAS. 
Second, the change patterns of the hand-developed models are mined to 
identify more suitable mutation operators for NAS.
	
We have realized \manas by extending \ak, the state-of-the-art NAS framework.
\ak is open source and
outperforms state-of-the-art methods like SEAS~\cite{elsken2018simple}, 
NASBOT~\cite{kandasamy2018neural} making it a suitable baseline~\cite{jin2019auto}. 
Some key technical contributions in \manas include {\em model matching}, a technique
for matching the problem that the developer intends to solve with the hand-developed 
models mined from repositories, 
{\em model adaptation}, a technique for adapting the mined model to the problem context, 
{\em model transformation}, 
a technique for adapting the mined model for further improving
metric values, and {\em training adaptation} that leverages 
mined parameter values from the repositories to change the optimizer. 
	
To evaluate \manas, we use the top-\nodata problems from diverse domains obtained from Kaggle for machine learning tasks including image classification and image regression. Our evaluation shows that given the same amount of searching time, \manas generates simpler neural architectures than \ak without losing accuracy. In terms of the models' size, \manas' models have {\bf 42.9\%} to {\bf 99.6\%} fewer numbers of parameters than \ak' models. We observed up to {\bf 641.6\%} faster training speed when training models 
produced by \manas as compared to those produced by \ak.
	
Our main contributions are the following:
\begin{itemize}
		\item We have proposed a novel approach for NAS that 
		leverages software repository mining.
		\item We have proposed methods to identify the suitable 
		models by analyzing data characteristics and adapting models.
		\item We have utilized the common patterns to transform 
		mined models to improve the performance of these models in 
		terms of error rate, MSE, model complexity, and training time.
		\item We have implemented these ideas in a SOTA NAS framework, 
		\ak~\cite{jin2019auto}. 
		{\bf Our artifact is available here~\cite{mnartifact}.} 
\end{itemize}

The paper is organized as follows: \S\ref{sec:motivation} presents a motivating example, \S\ref{sec:problem} presents preliminaries and problem statement, \S\ref{sec:manas} describes the \manas approach for NAS, \S\ref{sec:limit} describes the limitations and threats to validity of \manas,
\S\ref{sec:related} describes related work, and \S\ref{sec:conclude} concludes.

%

	\section{Motivation}
\label{sec:motivation}

Deep Learning has received much attention in both academia and industry. 
Therefore, many deep learning libraries and tools are created for supporting 
a large number of deep learning developers. 
Although these libraries and tools make deep learning more accessible, 
there are still many challenges. 
One of the challenging problems is constructing an appropriate CNN model architecture~\cite{wardat2021deepdiagnosis, wardat21deeplocalize}, 
which also has been shown as a frequent bug in CNN programs by Islam \etal ~\cite{islam19, islam20repairing}. 
For instance, \figref{fig:question} shows a query~\cite{so49226447} posted on \sof 
where a developer is unable to find an appropriate CNN model for their purpose. 
In particular, the question discusses the difficulty that the ResNet architecture does not 
give the result as the developer expected. In response, an expert suggests changing the CNN model. 
\figref{fig:answer} shows the solution of the expert for the question of the developer. 
The expert suggested that the developer should add dropout layers to minimize overfitting.

Neural architecture search (NAS) aims to solve this problem~\cite{DBLP:conf/iclr/ZophL17}.
NAS takes the training data as an input to automatically 
define the neural network for that data. 
Moreover, NAS is able to tune the hyperparameter 
of the searched neural network.
There are both commercial and open-source realizations of NAS. 
For example, a developer can pay about \$20 per 
hour to use Google's AutoML. 
\ak is an AutoML system using NAS~\cite{jin2019auto} 
created as an open-source alternative. 
\ak returns outstanding results compared with state-of-the-art handmade models 
on CIFAR10~\cite{krizhevsky2009learning}, 
MNIST~\cite{lecun1998gradient}, and 
FASHION~\cite{xiao2017fashion} datasets.
\ak is shown to outperform state-of-the-art methods like SEAS~\cite{elsken2018simple}, 
NASBOT~\cite{kandasamy2018neural}.

NAS has two limitations. 
First, it can be very costly. 
For example, \ak consumes 2,300\%~\cite{rosebrock2019ak} more GPU computation time 
compared to using handmade model.
Second, NAS often produces complex models that are hard to understand
and time-consuming to train.
To illustrate, we used \ak on another dataset, which is \emph{Blood Cell}~\cite{bloodcell2017kaggle} collected from \kg. 
The model created by \ak 
for the \emph{Blood Cell} problem has more than 2.3 million learnable parameters and more than 7 weight layers.
The searched CNN models are constructed based on the architecture of 
the large default CNN models; thus, the models produced by \ak are often really large.  
Smaller CNN models train faster and save more energy~\cite{iandola2016firecaffe, pan22decomposing}. 
Han \etal have shown that reducing the number of parameters of 
deep learning models can reduce the training time by 3$\times$ to 4$\times$, and energy comsumption by 3$\times$ to 7$\times$~\cite{han2015deep}. 
The rest of this work describes our approach \manas that addresses these limitations.
As an example, for \emph{Blood Cell} dataset, \manas produces a model that decreases
the error rate by 47.1\%, the model depth by 14.3\%, the model width by 87.0\%, and increases
the training speed by 56.9\% compared with \ak's model. 
The model produced by \manas has 6 layers and 0.3 million parameters (neurons),
whereas the model produced by \ak has 7 layers, 2.3 million parameters.

	 \section{Preliminaries and problem statement}
\label{sec:problem}

\begin{figure*}
	\includegraphics[width=1.0\textwidth]{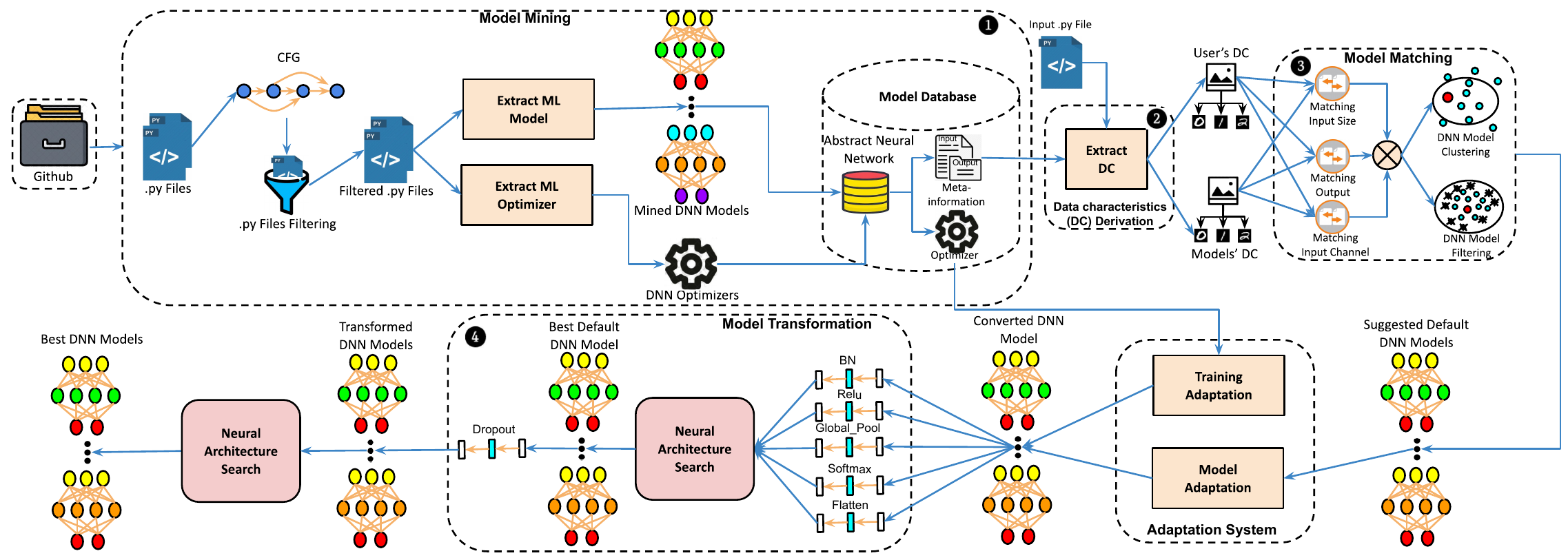}
	\caption{An Overview of \manas. Two inputs are mined models from repository (left-top), and user's initial file (middle-top). 
	}
	\label{fig:overview}
\end{figure*}

\subsection{Preliminaries (NAS)} 
We define NAS like it was defined in \ak~\cite{jin2019auto}. Given a search space $\mathcal{S}$ and the input data $\mathcal{D}$ split into $\mathcal{D}_{train}$ and $\mathcal{D}_{val}$, the goal is to find an optimal neural architecture  $\mathcal{N}$ $\in$ $\mathcal{S}$, which achieves the lowest value of the cost function in dataset $\mathcal{D}$. Search space $\mathcal{S}$ covers all the neural architectures created from default neural networks.
\begin{equation}
	\mathcal{N} = \underset{\mathcal{N} \in \mathcal{S}}{\arg \min}  \emph{Cost}(\mathcal{N'}(\theta'), \mathcal{D}_{val})
	\label{eq:1}
\end{equation}
\begin{equation}
	\theta' = \underset{\theta}{\arg \min}  \mathcal{L}(\mathcal{N'}(\theta), \mathcal{D}_{train})
	\label{eq:2}
\end{equation}
Where \emph{Cost} and $\mathcal{L}$ are the cost function and loss function, $\theta$ is the learned parameter of initial architectures $\mathcal{N'}$. 

\subsection{Problem Statement} This work aims to utilize the neural networks from open source repositories to optimize the neural architecture search. We extract the data characteristics from both the input dataset and mined neural networks to determine better starting points (initial models) for NAS. Instead of using concrete default neural networks, the goal is to find optimal initial architecture $\mathcal{N}^*$ for NAS for each different input dataset. Similar to Equation \ref{eq:2}, we obtain learned parameter $\theta^*$ of new  initial architectures $\mathcal{N}^*$:
\begin{equation}
	\theta^* = \underset{\theta}{\arg \min}  \mathcal{L}(\mathcal{N}^*(\theta), \mathcal{D}_{train})
	\label{eq:3}
\end{equation}

The optimal initial architectures $\mathcal{N}^*$ support NAS to find out the optimal neural network  faster. In other words, with the same amount of searching time and input dataset, new initial architectures help NAS to find out a neural network with lower error compared to the original NAS:
\begin{equation}
	{\arg \min}  \emph{Cost}(\mathcal{N}^*(\theta^*), \mathcal{D}_{val}) < {\arg \min}  \emph{Cost}(\mathcal{N'}(\theta'), \mathcal{D}_{val})
\end{equation}

\section{Manas}
\label{sec:manas}

Figure \ref{fig:overview} shows the overview of \manas. \manas has five major components
that we describe below. 

\begin{enumerate}
\item[\circled{1}] To initialize \manas for NAS, the model database must be populated by mining models from open source repositories. Currently, \manas collects high quality models from \gh by extracting Python files from \keras projects. These projects are selected using certain filtering criteria to ensure code quality. Then, API usage is used to filter Python files to those that contain models. Finally, both the models and the values for optimizer are extracted to store in the model database. This database is constructed once and should be updated periodically as new models are added frequently.   
\item[\circled{2}] The data characteristics extracted from both the input data and mined models are used to select the suitable initial models for NAS from the database.
\item[\circled{3}] Model matching matches the data characteristics extracted from an input dataset and the mined models to obtain a good starting point. It selects the models, which have the closest data characteristics with the input dataset by using the model clustering approach. In case there are too many initial models, the model filtering approach will be applied to reduce the number of models.
\item[\circled{4}] The selected models are transformed by the model transformation approach based on related state-of-the-art papers and common layer API patterns of mined models. The transformation can enhance the performance of the models in terms of errors and training speed.

\end{enumerate}

	\subsection{Model Mining}
\label{sec:mining}

\begin{figure}
	\includegraphics[width=0.47\textwidth]{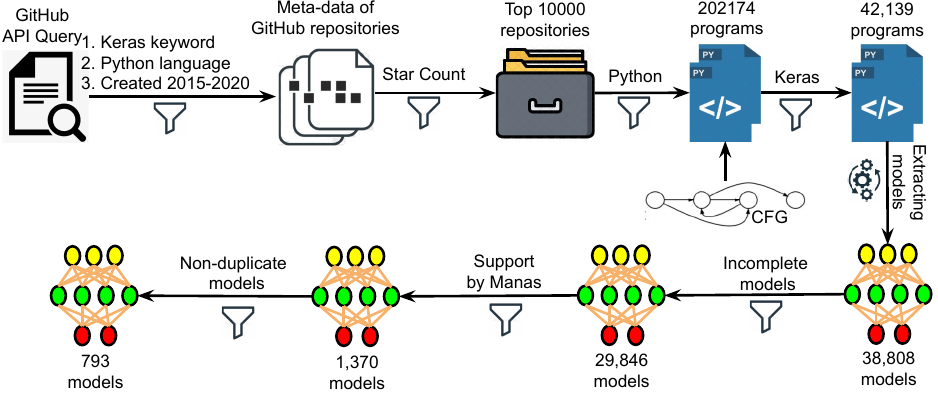}
	\caption{Model mining process.}
	\label{fig:miningoverview}
\end{figure}

In order to extract CNN models and their optimizers from source code repositories, 
we build a source code analyzer based on the control flow graph (CFG). Figure \ref{fig:miningoverview} shows the overview of model mining process.

\subsubsection{Meta-data collection} We collect the meta-data of \gh repositories by using a \gh API query ~\footnote{\url{"https://api.github.com/search/repositories?q=keras+language:python+created:yyyy-mm-dd"}}. Meta-data contains basic information of a repository such as authors, repository's name, etc. The query allows us to obtain the meta-data of the \gh repositories with three filtering criteria including Python programming language, containing \keras keyword, and created date between 2015-01-01 and 2020-12-31. From the meta-data, we obtain \gh URLs of the top 10,000 repositories with the most star count to ensure the quality of the models~\cite{borges2016understanding}. The URLs help us to access the repositories to collect the CNN models.

\subsubsection{Keras programs collection}
We obtain 202,174 Python programs from collected repositories; however, only 42,139 programs use \keras API. In particular, we use CFG to analyze their import statements of Python programs to only collect which one imports \keras API.


\subsubsection{Models extraction} In this work, since \manas only works with \keras CNN, we will explain how we extract CNN from \keras programs.  We use CFG to extract a model from a deep learning program. We manually create a list of function calls used to build neural networks of \keras based on \keras' documentation~\cite{kerasapi}. Then, CFG examines every API call to collect the functions contained in the list and their connections. The collected functions represent the layers in the models. The connections between functions represent the layer connections in the models. The reason for using CFG is to collect complete models from programs even if they contain branches. For example, the CFG contains a convolutional block followed by a dense block in the "if" branch and a skip connection in the "else" branch. If a convolutional block combines with a dense block or a convolutional block combines with a skip connection to be a complete model, we will extract those parts in the branches to collect the complete model. CFG is used to handle the situation that there is a loop containing a part of a model in the CNN programs. If the number of iterations is available to extract that part completely, we will collect it to complete the model. If there is a method call in the program containing a part of the model, we will collect it to complete the model. For the other cases, when the parts in the branches cannot complete the model or cannot be extracted, we will ignore them. For instance, a part of the model in a loop whose numbers of iterations are unavailable cannot be extracted. After this step, we collected 38,808 models from 42,139 \keras programs. The number of models is less than the number of \keras programs because many programs do not contain models. We assume that a program contains a model if it has at least an API used to build a model.

Rather than mining only neural architectures, we also mine their optimizers that deep learning developers carefully select after spending manual efforts on retraining their models. Then, whenever a model is selected as an initial model, \manas trains the model with its optimizer instead of the default one. Optimizers are algorithms deciding how the parameters of the models change. Every optimizer has strengths and weaknesses; thus, it is necessary to choose a suitable optimizer. While models are the decisive factor to the performance of \manas, we note that it is wasteful if we cannot fully use these models. In other words, good models with wrong optimizers cannot produce a good performance. To obtain the optimizer, we use the same process of extracting CNN models by creating a list of functions related to the optimizer. After that, CFG analysis is used to obtain the API call, which contains the optimizer. 

\subsubsection{Incomplete models detection}
A complete model includes an input layer, hidden layers, and an output layer~\cite{ZHU2020125}. The input layer is the first convolutional layer of the neural network, which is distinguished from the other convolutional layers based on its parameters. In particular, only the API call representing the input layer contains \emph{input_shape} parameter. The output layer is the last linear layer of the CNN. Hidden layers are the layers between the input layer and the output layer, including convolutional layers, activation layers, and fully connected layers (linear layers). Therefore, if an extracted model does not have convolutional layers, activation layers, or linear layers, we will consider that the model is incomplete. By removing incomplete models, there are 29,846 models left.

\subsubsection{Supported by Manas}
As \manas currently supports a few kinds of layers that are the convolutional layer, the linear layer, the batch normalization layer,
the concatenate layer, the add layer, the max pooling layer, the dropout
layer, the Softmax layer, the ReLU layer, the flatten layer, and the global
pooling layer, we filter out the models containing unsupported layers. After this step, we have 1,370 models left.

\subsubsection{Model duplication detection}
We obtain 793 models after removing the duplicate ones. 
We consider if two models have the same 
abstract neural network (ANN), there will be a duplicate model. We store the extracted CNN models in a database as an abstract neural network, 
which is an abstract representation of the neural network. 
This representation has the structure as a network where the nodes are API calls, 
and the connections are the order between API calls. 
We use this representation to adapt models and their optimizers into \manas. From each node, we obtain the name of layer and its parameters, which can be converted into an API call. 
Figure \ref{fig:converted} presents an example of ANN built from an mined model. Notice that
if an activation parameter is implemented as an argument inside a layer, 
we consider that the activation is a separate layer.
\begin{figure}
	\centering
	\begin{subfigure}[b]{.45\linewidth}
		\begin{lstlisting}[basicstyle=\tiny,numberblanklines=false]
		Conv2D(64, kernel_size = (3, 3),activation='relu',input_shape=(3, 120, 180))
		
		MaxPooling2D(pool_size=(2,2))
		Conv2D(32,                   activation='relu')
		
		MaxPooling2D(pool_size=(2,2))
		Dropout(0.25)
		Flatten()
		Dense(20, activation='softmax')


		SGD(lr=0.01, decay=1e-6) @\label{line:optim}@
\end{lstlisting}
		\caption{Extracted model}
		\label{fig:originalCNN}
	\end{subfigure}
	\begin{subfigure}[b]{.45\linewidth}
		\begin{lstlisting}[basicstyle=\tiny,numberblanklines=false]	
		{'func': 'Conv2D', 'input_shape': [3, 120, 180], 'arg2': 64, 'kernel_size': [3, 3]} @\label{line:example}@
		{'func': 'relu'}
		{'func': 'MaxPool2d', 'pool_size': [2, 2]}
		{'func': 'Conv2D', 'arg1': 64, 'arg2': 32}
		{'func': 'relu'}
		{'func': 'MaxPool2d', 'pool_size': [2, 2]}
		{'func': 'Dropout', 'arg1': 0.25}
		{'func': 'Flatten'}
		{'func': 'linear', 'arg1': 128, 'arg2': 20}
		{'func': 'softmax'}
		
		{'func': 'SGD', 'lr': 0.01, 'decay': 1e-06} @\label{line:optimconvert}@
\end{lstlisting}
		\caption{Abstract Neural Network}
		\label{fig:convertedCNN}
	\end{subfigure}
	\caption{Building an abstraction of neural network from extracted model}
	\vspace{-5mm}
	\label{fig:converted}
\end{figure}

	\subsection{Data Characteristics Derivation}
\label{sec:intent}

We  analyzes the input data and input/output layers of the mined models to extract their data characteristics. Taking an image classification as an example, we obtain the input data characteristics including the input size and the output channel from the image data.  The data characteristics extracted from a mined model also are the input size and the output channel are obtained by analyzing the first convolutional layer and the last linear layer of a CNN model, which are the input layer and the output layer the model, respectively.

\begin{example}
Line \ref{line:inputlayer} is an API call representing for the input layer of a CNN. We extract the value of the argument \emph{input\_shape} from the input layer to obtain the input size and the input channel, which are (120, 180) and 3, respectively. 
The output channel is obtained from the output layer \ref{line:outputlayer}. By extracting the first argument's value of the output layer, we obtain the value of the output channel.

\begin{lstlisting}[language=Python,columns=fullflexible,numbers=none]
Conv2D('input_shape': [3, 120, 180], activation='relu')@\label{line:inputlayer}@
...
Dense(10, activation='relu')@\label{line:outputlayer}@
\end{lstlisting}
\end{example}

Since we currently focus on image problems like image classification and image regression, we use the input size, the input channel, and the output channel of image as data characteristics. First, the input size includes the height and the width extracted from the input data and mined models. Second, the input channel represents the number of primary colors in the image. Third, the output channel is the number of output categories of the data and models.

	\subsection{Model Matching}
Model matching is a ranking system used to find good models for a certain 
problem by using the data characteristics. Instead of using constant default models, model matching finds the suitable models to uses them as new default models for NAS.
\label{sec:matching}

\subsubsection{Model clustering} Model clustering uses the data characteristics of both the input dataset and mined neural networks to select appropriate initial architectures $\mathcal{N}^*$ for NAS. First of all, \manas clusters the mined models based on the data characteristics of the models and the input dataset. Secondly, in a meta-feature (data characteristics) space, our approach identifies closest clusters to the input dataset. Lastly, \manas uses all the models in the closest cluster to the dataset as the initial architectures for NAS. Formally, we determine the initial architectures $\mathcal{N}^*$ for dataset $\mathcal{D}$ in Equation \ref{eq:3} as follows: 
\begin{equation}
	\mathcal{N}^* = \{\mathcal{C}_k \:|\: n \in C_k, \mathcal{C}_k \in \mathcal{C}\}
	\label{eq:initmodels}
\end{equation}
\begin{equation}
	\mathcal{C} = \{\mathcal{G}(\Delta o_k, \Delta s_k) \: | \:	 \Delta i_k = 0, k \in [1, |\mathcal{S}|]\} 
 \label{eq:gmeans}
\end{equation}
\begin{equation}
\emph{dist}(n, \mathcal{D}) = 	
	\begin{cases} 
		{\min} \; \Delta o \; \textnormal{if} \; {\min} \;\Delta i = 0 \\
		{\min} \; \Delta s \; \textnormal{if} \; {\min} \;\Delta i = 0, {\min} \;\Delta o = 0
	\end{cases}
    \label{eq:distance}
\end{equation}
In Equation \ref{eq:gmeans}, $\mathcal{C}$ is a set of clusters of neural network detected by clustering algorithm G-means~\cite{hamerly2004learning}, which uses a statistical test to automatically decide the number of clusters. $\Delta o_k$,  $\Delta i_k$, and $\Delta s_k$ are measured as follows:
\begin{equation}
	\Delta i_k = |i - i_k| \label{eq:6}
\end{equation}
\begin{equation}
	\Delta o_k = |o - o_k| \label{eq:4}
\end{equation}
\begin{equation}
	\Delta s_k = \sqrt{(w - w_k)^2 + (h - h_k)^2}  \label{eq:5}
\end{equation}	
Where $i$, $o$, $w$, and $h$ are the input channel, the number of output classes, the input width, and the input height of the dataset, respectively. Similarly, $i_k$, $o_k$, $w_k$, and $h_k$ are the input channel, the number of output classes, the input width, and the input height of a model k, respectively. The idea behind the clustering equations is that there are two types of the input channel, including 1 and 3. Therefore, we classify the neural networks that have the same input channel first. Then, we use ($\Delta o$, $\Delta s$) as the input of G-means to split the mined models into clusters. After that we identify the closest model $n$ to the input dataset like Equation \ref{eq:distance}. The closest model is identified based on the priority of $\Delta o$ and $\Delta s$ that $\Delta o$ takes precedence over $\Delta s$. We have tried to run our tool in different orders of priority $\Delta o$ and $\Delta s$; however, this order of priority gives us the best results. The closest cluster to the input dataset is the cluster which contains the closest model. Then, we select all the models in the closest cluster to the input dataset as the initial models $\mathcal{N}^*$ for NAS shown in Equation \ref{eq:initmodels}. 

\subsubsection{Model filtering} If the number of initial models found by the model clustering approach is too large, we use model filtering to filter some models to increase the performance of \manas. In the searching process, \manas trains all the default neural networks to select the best one. After that NAS is applied to tune the selected model. Thus, with a specific time budget, the more time \manas spends on trying the default models, the less time it spends on NAS for model searching. To filtering neural networks, we detect the equivalent architectures. We treat each neural architecture as a graph, whose trainable layers like convolutional layers or dense layers represent vertices, connections between two trainable layers represent edges, and channels of the outgoing trainable layers represent the weights of edges between two vertices. We use Cosine similarity to measure the similarity between two vertices. $M_{ij}$ is an element of the similarity matrix of vertex $v_i$ of a graph $G_A$ and vertex $v_j$ of a graph $G_B$, which is measured as follows:
\begin{equation}
	M_{ij} = \frac{ \sum_{k=1}^{n} w_{ik}w_{jk} }{\sqrt{\sum_{l=1}^{m} w_{il}^2}\sqrt{\sum_{n=1}^{k} w_{jl}^2}}
\end{equation}
Suppose that $v_k$ is the common neighbor of vertex $v_i$ and vertex $v_j$, we have $w_{ik}$ and $w_{jk}$ are the weights of edges $v_kv_i$ and $v_kv_j$, respectively. We also have $w_{il}$ or $w_{jl}$ are the weights of edges that incident with each of the vertices $v_i$ and $v_j$, respectively.
\begin{equation}
 w_{pq} = 	
 \begin{cases} 
 \textnormal{weight of the edge between $v_{p}$ and $v_{q}$ in a graph $G$} \\
 0 \textnormal{ otherwise}	
 \end{cases}
 \label{eq:adjmatrix}
\end{equation}
We create a similarity matrix $M$ for each pair of graphs and classify equal matrices into the same classes. In this work, we filter the models by only using the neural network whose graphs belong to the class having the highest number of matrices. In other words, we will only choose the most used architecture in the determined initial architecture for NAS. We consider the most frequent architecture to be more suitable than others because many deep learning developers repeatedly use these neural architectures. Non-trainable layers like dropout or activation are not used as the factors to detect the similarities of the neural architectures because we want to utilize the various usages of these layers to increase the performance of \manas. Particularly, different models can have the same graph structure because of the differences in non-trainable layers; therefore, a graph can refer to many different neural architectures.

	\subsection{Model Transformation}
Even though the mined models are from the top star \gh repositories, they are not perfect. 
Therefore, we transform the initial architectures by modifying or adding the batch normalization (BN) layer, the flatten layer, the activation layers, the global average pooling (GAP) layer, and the dropout layer to optimize NAS in terms of speed, errors, and the number of parameters. We choose these layers to modify the default models because of two reasons. Firstly, we have found many common patterns related to these layers. Secondly, many recent studies have shown the effectiveness of these layers on increasing the performance of deep learning models. We have created a set of pre-defined rules to transform networks based on related state-of-the-art papers and common function calls' patterns from mined models. \manas analyzes these architectures to determine whether the network satisfies the pre-defined rules. If these rules are satisfied, the pre-defined model transformations will be applied. The pre-defined model transformations support NAS to ignore transformations offered by our approach and focus on the other transformations. The model transformations do not work for \ak' initial neural architectures because those default models have already included these transformations.
\begin{figure}[ht]
	\centering
	\begin{subfigure}[b]{.45\linewidth}
		\begin{lstlisting}[basicstyle=\tiny,numberblanklines=false]
		(0): Conv2d(3, 32, ...)
		
		(1): Tanh()
		(2): MaxPool2d(kernel=2, stride=2, ...)
		
		(3): Conv2d(32, 32, ...)
		
		(4): Tanh()
		(5): MaxPool2d(kernel=2, stride=2, ...)
		
		(6): Flatten()
		(7): Linear(in=32, out=32, ...)
		(8): ReLU()
		
		(9): Linear(in=32, out=2, ...)
		(10): Softmax()
\end{lstlisting}
		\caption{Original model}
		\label{fig:original}
		
\end{subfigure}
	\begin{subfigure}[b]{.45\linewidth}
		\begin{lstlisting}[basicstyle=\tiny,numberblanklines=false]
		(0): Conv2d(3, 32, ...)
		(1): BatchNorm2d(32, ...) @\label{line:BN1}@
		(2): ReLU()
		(3): MaxPool2d(kernel=2, stride=2, ...)
		(4): Dropout2d(p=0.5) @\label{line:drop1}@
		(5): Conv2d(32, 32, ...)
		(6): BatchNorm2d(32, ...) @\label{line:BN2}@
		(7): ReLU()
		(8): MaxPool2d(kernel=2, stride=2, ...)
		(9): Dropout2d(p=0.5) @\label{line:drop2}@
		(10): GlobalAvgPool2d() @\label{line:average}@
		(11): Linear(in=32, out=32, ...)
		(12): ReLU()
		(13): Dropout2d(p=0.25) @\label{line:drop3}@
		(14): Linear(in=32, out=2, ...)
		(15): Softmax() @\label{line:activation}@
\end{lstlisting}
		\caption{Transformed model}
		\label{fig:transformed}
		
\end{subfigure}
	\caption{Original CNN model vs transformed CNN model}
	\label{fig:transvsorigin}
\end{figure}

\subsubsection{Batch normalization layer constraint} We add a new batch normalization layer between the convolutional layer and the activation layer to increase training speed~\cite{ioffe2015batch}. 
Using BN means that we modify the activations to normalize the input layer to decrease the training time. Many well-known neural architectures like ResNet~\cite{he2016deep}, DenseNet~\cite{huang2017densely}, EfficientNet~\cite{tan2019efficientnet} use BN to increase the training speed. BN is also popular in optimizing NAS \cite{chen2019nips, elsken2018efficient, DBLP:conf/iclr/ZophL17}.
\begin{example}
	According to Figure \ref{fig:original}, the original model has a convolutional layer connects to a ReLU layer, 
	which is an activation function layer. Thus, in Figure \ref{fig:transformed}, 
	following the batch normalization layer constraint, \manas adds a new BN layer between the 
	convolutional layer and the ReLU layer like the following example.
	\vspace{-1mm}
	\begin{figure}[H]
		\centering
		\begin{subfigure}[b]{.45\linewidth}
			\begin{lstlisting}[basicstyle=\tiny,numberblanklines=false]
		(0): Conv2d(3, 32, ...)

		(1): ReLU()
\end{lstlisting}
\caption{Original model}
		\end{subfigure}
		\begin{subfigure}[b]{.45\linewidth}
			\begin{lstlisting}[basicstyle=\tiny,numberblanklines=false]	
		(0): Conv2d(3, 32, ...)
		(1): BatchNorm2d(32, ...) 
		(2): ReLU()
\end{lstlisting}
\caption{Transformed model}
		\end{subfigure}
	\caption{Example of batch normalization layer constraint}
	\label{fig:bn}
	\end{figure}
\end{example}	
\subsubsection{Global average pooling layer constraint} We use GAP to reshape the data into the correct format for fully connected layers to prevent overfitting~\cite{lin2013network}. Moreover, we use the mined models as default models for NAS; thus, the
original input size of the initial models and the input size of the dataset can be different, which can cause a shape mismatch bug. However, using GAP can solve this problem since it does not care about the input shape. 
\begin{example}
	According to Figure \ref{fig:original}, the original model used the flatten layer to pass the feature map through the CNN. Therefore, in Figure \ref{fig:transformed}, following the constraint about GAP layer, \manas transforms flatten into GAP like the following example.
	
	\begin{figure}[H]
	\centering
	\begin{subfigure}[b]{.45\linewidth}
		\begin{lstlisting}[basicstyle=\tiny,numberblanklines=false]
		Flatten()
		Linear(in=32, out=32, ...)
\end{lstlisting}
\caption{Original model}
	\end{subfigure}
	\begin{subfigure}[b]{.45\linewidth}
		\begin{lstlisting}[basicstyle=\tiny,numberblanklines=false]	
		GlobalAvgPool2d()
		Linear(in=32, out=32, ...)
\end{lstlisting}
\caption{Transformed model}
	\end{subfigure}
	\caption{Example of global average pooling layer constraint}
\end{figure}
\end{example}	
\subsubsection{Activation layer constraint} We investigate the patterns of usages of activation functions used in the mined model. We have found 3218 hidden layers used in 793 models, where ReLU is used 2946 times, accounting for 94.55\%. Therefore, we replace the current activation layers of convolutional layers with ReLU.
\begin{example}
    According to Figure \ref{fig:original}, the original model uses Tanh for the convolutional layer. Thus, following the constraint of the activation layer, \manas transforms Tanh to ReLU like following example like the following example.
	\begin{figure}[H]
		\centering
		\begin{subfigure}[b]{.45\linewidth}
	\begin{lstlisting}[basicstyle=\tiny,numberblanklines=false]
	(3): Conv2d(32, 32, ...)
	
	(4): Tanh()
\end{lstlisting}
\caption{Original model}
\end{subfigure}
\begin{subfigure}[b]{.45\linewidth}
	\begin{lstlisting}[basicstyle=\tiny,numberblanklines=false]
	(5): Conv2d(32, 32, ...)
	(6): BatchNorm2d(32, ...) 
	(7): ReLU()
\end{lstlisting}
\caption{Transformed model}
\end{subfigure}
\caption{Example of activation layer constraint}
\end{figure}
\end{example}
\subsubsection{Dropout layer constraint} We investigate the frequency of dropout~\cite{srivastava2014dropout} layers and their drop rates used in the mined model.
Out of 806 times that dropout is used in hidden layers, the drop rate of 0.25 is used 385 times, which
accounts for 47.8\%. 
Out of 753 times that dropout is used in fully connected layers, the drop rate of 0.5 is used 529 times, which
accounts for 70.3\%. 
Thus, we add dropout layers with a drop rate of 0.25 and 0.5 to the hidden layers and fully connected layers, respectively.
\begin{example}
	According to Figure \ref{fig:original}, the original model does not use the dropout layer. Therefore, in Figure \ref{fig:transformed}, following dropout layer constraint, \manas adds dropout layers into a convolutional layer with 0.25 drop rate and a dropout layer into the fully connected layer with 0.5 drop rate like the following example.
	\begin{figure}[H]
	\centering
	\begin{subfigure}[b]{.45\linewidth}
		\begin{lstlisting}[basicstyle=\tiny,numberblanklines=false]
	Conv2d(32, 32, ...)
	ReLU()
	MaxPool2d(kernel=2, stride=2, ...)
	
\end{lstlisting}
\caption{Original model}
	\end{subfigure}
	\begin{subfigure}[b]{.45\linewidth}
		\begin{lstlisting}[basicstyle=\tiny,numberblanklines=false]	
	Conv2d(32, 32, ...)
	ReLU()
	MaxPool2d(kernel=2, stride=2, ...)
	Dropout2d(p=0.5)
\end{lstlisting}
\caption{Transformed model}
	\end{subfigure}
\caption{Example of dropout layer constraint}
\end{figure}
\end{example}
All the transformations except dropout transformation are applied simultaneously to the mined models before searching. After the best model is selected from candidate models, dropout transformation is applied to the best model. Since adding dropout does not always improve the model's errors, we utilize NAS to identify whether using dropout is good or bad.
	\section{Evaluation}

\begin{table*}[ht]
	\centering
	\caption{\manas Classification \& Regression Results}
 	\setlength{\tabcolsep}{1.2pt}
	\footnotesize
\begin{tabular}{|c|r|r|r|r|r|r|r|r|r|r|r|r|r|r|r|r|}\hline
\multirow{3}{*}{Data} & \multicolumn{8}{c|}{Image Classification}                                                                                                                                                                                                                                                                                             & \multicolumn{8}{c|}{Image Regression}                                                                                                                                                                                                                                                  \\ \cline{2-17} 
                     & \multicolumn{2}{c|}{\begin{tabular}[c]{@{}c@{}}Error rate\\ (\%)\end{tabular}} & \multicolumn{2}{c|}{\begin{tabular}[c]{@{}c@{}}Depth\\ (layers)\end{tabular}} & \multicolumn{2}{c|}{\begin{tabular}[c]{@{}c@{}}Param \#\\ (million)\end{tabular}} & \multicolumn{2}{c|}{\begin{tabular}[c]{@{}c@{}}Speed\\ (epoch/min)\end{tabular}} & \multicolumn{2}{c|}{MSE} & \multicolumn{2}{c|}{\begin{tabular}[c]{@{}c@{}}Depth\\ (layers)\end{tabular}} & \multicolumn{2}{c|}{\begin{tabular}[c]{@{}c@{}}Param \#\\ (million)\end{tabular}} & \multicolumn{2}{c|}{\begin{tabular}[c]{@{}c@{}}Speed\\ (epoch/min)\end{tabular}} \\ \cline{2-17}
                      & \multicolumn{1}{c|}{AK}                                     & \multicolumn{1}{c|}{MN}                                    & \multicolumn{1}{c|}{AK}                                     & \multicolumn{1}{c|}{MN}                                   & \multicolumn{1}{c|}{AK}                                       & \multicolumn{1}{c|}{MN}                                     & \multicolumn{1}{c|}{AK}                                      & \multicolumn{1}{c|}{MN}                                     & \multicolumn{1}{c|}{AK}             & \multicolumn{1}{c|}{MN}             & \multicolumn{1}{c|}{AK}                                     & \multicolumn{1}{c|}{MN}                                   & \multicolumn{1}{c|}{AK}                                      & \multicolumn{1}{c|}{MN}                                      & \multicolumn{1}{c|}{AK}                                      & \multicolumn{1}{c|}{MN}                                     \\ \hline
Blood Cell            & 18.9                                   & 10.0 ($\downarrow${
 47.1\%})                                   & 7                                      & 6 ($\downarrow${
 14.3\%})                                    & 2.3                                      & 0.3 ($\downarrow${
 87.0\%})                                      & 5.1                                     & 8.0 ($\uparrow${
 56.9\%})                                   & 0.16           & 0.11 (\bf $\downarrow${
31.3 \%})            & 21                                     & 9 ($\downarrow${
 57.1\%})                                   & 11.2                                     & 0.1 ($\downarrow${
 99.1\%})                                      & 2.7                                     & 8.4 ($\uparrow${
 211.1\%})                                 \\ \hline
Breast Cancer         & 6.9                                    & 6.9 ($\downarrow${
 0.0\%})                                  & 21                                     & 8 ($\downarrow${
 61.9\%})                                   & 11.2                                     & 0.5 ($\downarrow${
 95.5\%})                                   & 0.2                                     & 0.5 ($\uparrow${
 150\%})                                  & 0.05           & 0.06 ($\uparrow${
 20.0\%})           & 26                                     & 7 ($\downarrow${
 73.1\%})                                    & 11.4                                    & 0.04 ($\downarrow$\bf 99.6\%)                                     & 0.2                                     & 0.6 ($\uparrow${
 200.0\%})                                  \\ \hline
Flower                & 14.6                                   & 12.8 ($\downarrow${
 12.3\%})                                     & 121                                    & 11 ($\downarrow${
\bf 90.9\%})                                   & 7.0                                      & 1.5 ($\uparrow${
 78.7\%})                                    & 2.7                                     & 3.8 ($\uparrow${
 40.7\%})                                 & 0.64           & 0.65 ($\downarrow${
 1.5\%})           & 21                                     & 9 ($\downarrow${
 57.1\%})                                   & 11.2                                    & 0.4 ($\downarrow${
 96.4\%})                                     & 3.3                                     & 4.3 ($\uparrow${
30.3\%})                                   \\ \hline
IIC                   & 8.9                                    & 8.6 ($\downarrow${
3.4\%})                                  & 27                                     & 7 ($\downarrow${
74.1\%})                                   & 15.5                                     & 0.5 ($\downarrow${
96.8\%})                                     & 0.7                                     & 2.6 ($\uparrow${
73.1\%})                                 & 0.58           & 0.59  ($\uparrow${
1.7\%})          & 122                                    & 5 (\bf$\downarrow$ 95.9\%)                                    & 7.0                                     & 0.3 ($\downarrow$ 95.7\%)                                   & 0.7                                     & 2.4 ($\uparrow$ 242.9\%)                                  \\ \hline
Malaria               & 2.3                                    & 1.6 ($\downarrow$ 30.4\%)                                   & 30                                     & 7 ($\downarrow$ 76.7\%)                                   & 20.3                                     & 0.3 (\bf $\downarrow$ 98.5\%)                                    & 1.7                                     & 4.3 ($\uparrow$ 152.9\%)                                    & 0.02           & 0.02 ($\downarrow$ 0\%)           & 21                                     & 8 ($\downarrow$ 61.9\%)                                    & 11.2                                    & 0.4 ($\downarrow$ 96.4\%)                                    & 2.4                                     & 4.7 ($\uparrow$ 95.8\%)	                                   \\ \hline
MNIST: Ham            & 20.6                                   & 20.3 ($\downarrow$ 1.5\%)                                 & 7                                      & 6 ($\downarrow$ 14.3\%)                                     & 0.9                                      & 1.6 ($\uparrow$ 77.8\%)                                  & 2.2                                     & 1.4 ($\downarrow$ 36.4\%)                                 & 0.81           & 0.73 ($\downarrow$ 9.9\%)            & 26                                     & 13 ($\downarrow$ 50.0\%)                                  & 11.9                                    & 0.4 ($\downarrow$ 96.6\%)                                    & 1.1                                     & 5.2 (\bf \textbf{$\uparrow$ 372.7\%})                                   \\ \hline
SD                    & 0.5                                    & 0.0 (\bf$\downarrow$ 100.0\%)                                   & 22                                     & 11 ($\downarrow$ 50.0\%)                                   & 11.3                                     & 0.7 ($\downarrow$ 93.8\%)                                      & 8.9                                    & 66.0   ($\uparrow$\bf 641.6\%)                                   & 0.05           & 0.1 ($\uparrow$ 100.0\%)            & 21                                     & 12 ($\downarrow$ 42.9\%)                                   & 11.2                                    & 0.4 ($\downarrow$ 96.4\%)                                     & 48.4                                   & 78.2  ($\uparrow$ 61.6\%)                                  \\ \hline
SL                    & 0.0                                    & 0.0  ($\downarrow$ 0.0\%)                                 & 22                                     & 6  ($\downarrow$ 72.7\%)                                  & 11.3                                      & 0.3  ($\downarrow$ 97.3\%)                                    & 5.5                                     & 8.0 ($\uparrow$ 45.5\%)                                    & 0              & 0 ($\downarrow$ 0.0\%)              & 27                                     & 10 ($\downarrow$ 63.0\%)                                   & 12.6                                    & 7.2 ($\downarrow$ 42.90\%)                                      & 5.0                                     & 2.6 ($\downarrow$ 48.0\%)                                     \\ \hline \hline
Avg                   & 9.1                                   & 7.5 ($\downarrow$\bf\errorrate)                                  & 32.1                                 
& 7.8 ($\downarrow$\bf\depth)                                & 10.0                                    & 0.7 ($\downarrow$\bf\width)                                    & 3.4                                   & 11.8 ($\uparrow$\bf\speed)                                    & 0.3           & 0.3 ($\downarrow$\bf\mse)           & 35.6                                   & 9.1 ($\downarrow$\bf\depthreg)                                   & 11.0                                    & 1.2 ($\downarrow$\bf\widthreg)                                   & 8.0                                   & 13.3 ($\uparrow$\bf\speedreg)                                  \\ \hline
\end{tabular}
\label{tbl:results}
\centering

\scriptsize \* In the table, Avg, AK and MN represent average, \ak, and \manas, respectively. In each cell, $\downarrow$ and $\uparrow$ represent a decrease percentage and an increase percentage, respectively.

In each evaluation metric, the bold value show the best percentage change of \manas compared to \ak. 
\vspace{-3mm}
\end{table*}

%
%
%

\label{sec:experiments}
%

%
%
%
%

%

\subsection{Experimental Setup} 
We implement \manas by extending \ak~\cite{jin2019auto}. 
All experiments use Python 3.6, with 16GB GPU Tesla V100. 
In these experiments, we use \nodata datasets for image classification and image regression, 
which are obtained from \kg based on the vote count. 
The efficiency and effectiveness of \manas are evaluated in three aspects. 
Firstly, we evaluate the metric values including error rate and MSE that is lower the better, model complexity, and training speed of \manas by comparing it with \ak.  
Secondly, we evaluate the model matching approach's efficiency and effectiveness. 
Lastly, we evaluate the efficiency and effectiveness of model transformation and training 
adaptation approaches. In these comparisons, \manas include model matching, model transformation, and NAS. The algorithm NAS used by \manas explores the search space via morphing the neural architectures guided by the Bayesian optimization algorithm. Mining the models is the most time-consuming task, which took about 24 hours to complete mining all the models. However, we only do this one time, so we do not count the time to mine the model in the comparison of \manas and \ak.

\subsection{Mined Models} 
Since many models for different \kg datasets have already been published on \gh, it is possible that some
\kg models of the testing dataset have already been included in our mined models. To avoid this problem, for each
	dataset, we have examined all \gh repositories of selected initial models of each dataset. There is no information showing that those initial models are specifically created for the input dataset. Moreover, for each input dataset, we have compared the initial model of \manas with all models from \kg. \manas’ initial model is not one of \kg’s models. Moreover, since most of
	the models from \kg are published as Jupyter Notebooks,
	we only mine the model written as Python files.

We only collect models from the top 10,000 repositories with the most star count to ensure high-quality models. We also do sanity checks by removing incomplete models and duplicate models. Moreover, we evaluate the mined models by training them on MNIST with 50 epochs~\cite{pan22decomposing}. The average accuracy of the mined models on MNIST is 90.98\%.

\subsection{Datasets} 
To evaluate the performance of our method, we use {\bf \nodata} different image 
datasets collected from \kg:   
\emph{Blood Cell}~\cite{bloodcell2017kaggle}, 
\emph{Breast Cancer}~\cite{breast2017kaggle},  
\emph{Flower}~\cite{flower2017kaggle}, 
\emph{Intel Image Classification} (IIC)~\cite{IIC2018kaggle},
\emph{Malaria}~\cite{malaria2018kaggle}, 
\emph{MNIST: Ham}~\cite{ham2018kaggle}, 
\emph{Sign Language Digits} (SD)~\cite{sd2017kaggle}, 
and \emph{Sign Language} (SL)~\cite{sl2017kaggle}. 
Our goal was to utilize more complex datasets compared to
well-known datasets such as MNIST, CIFAR10, and FASHION. 
Most of our datasets have large sizes of
images. For example, in the intel image classification~\cite{IIC2018kaggle}
dataset, the image size is 150x150 while the image size of
MNIST is only 28x28. Secondly, the number of images in our datasets 
is much larger than MNIST, CIFAR10, and FASHION. 
For instance, the breast cancer dataset~\cite{breast2017kaggle}
has 277,524 images. Lastly, the number of classification of
the evaluated datasets is up to 10 classes. These dataset are often used for image classification task; however, we treat prediction targets as numerical values for image regression task. For example, we will treat the prediction targets of the MNIST dataset as integers ranging from 0 to 9 as numerical values to be directly used as the regression targets. We only use image datasets because \ak only apply NAS for image data. we divide it into two subsets, 80\% of randomly selected images are used for training and the remaining images for validation.

\begin{figure*}[]
\begin{subfigure}[b]{0.24\textwidth}
	\includegraphics[keepaspectratio = True, scale = 0.29]{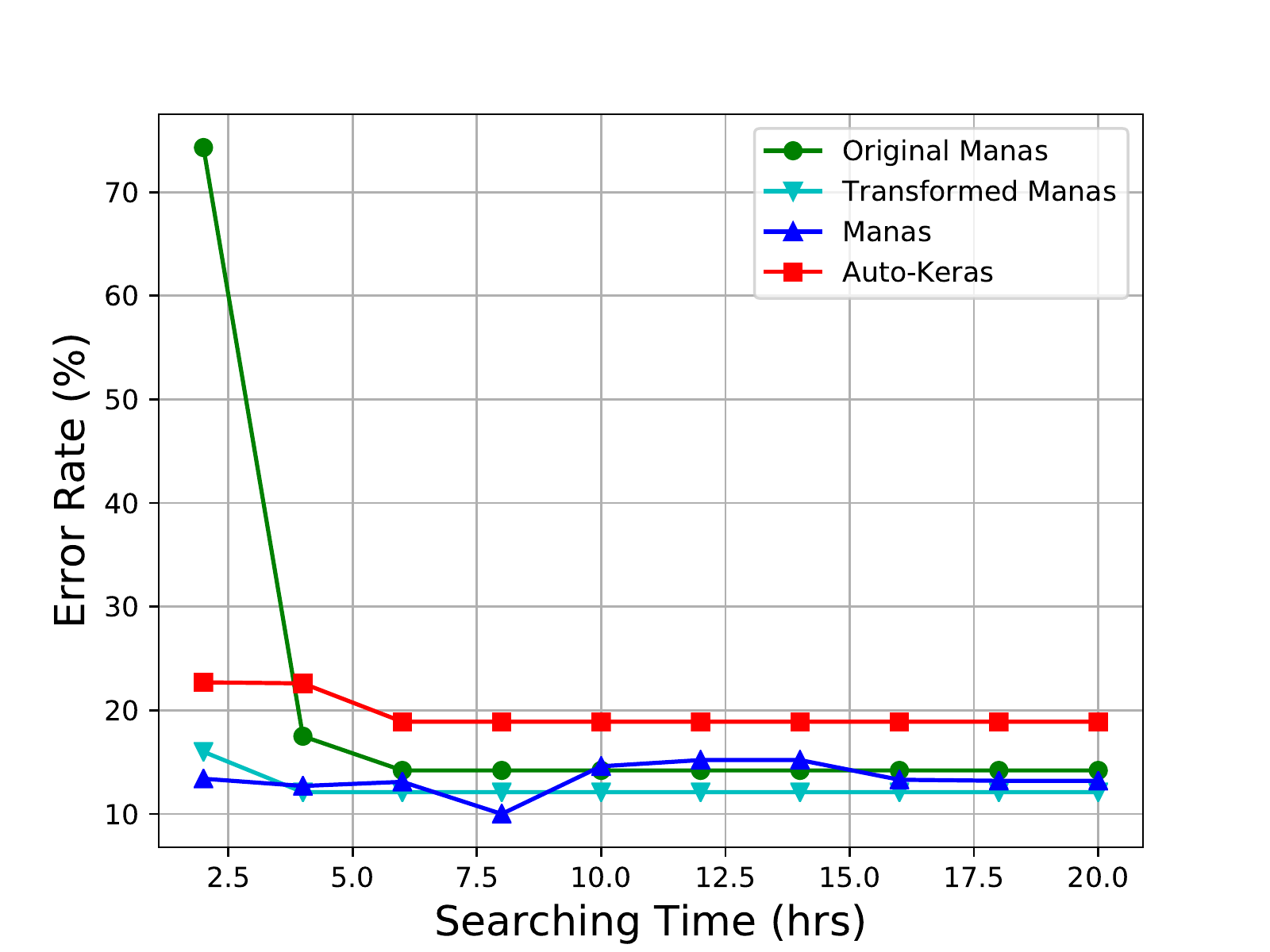}
	\caption{Blood Cell Classification}
	\label{fig:bc}
\end{subfigure} 
\begin{subfigure}[b]{0.24\textwidth}
	\includegraphics[keepaspectratio = True, scale = 0.29]{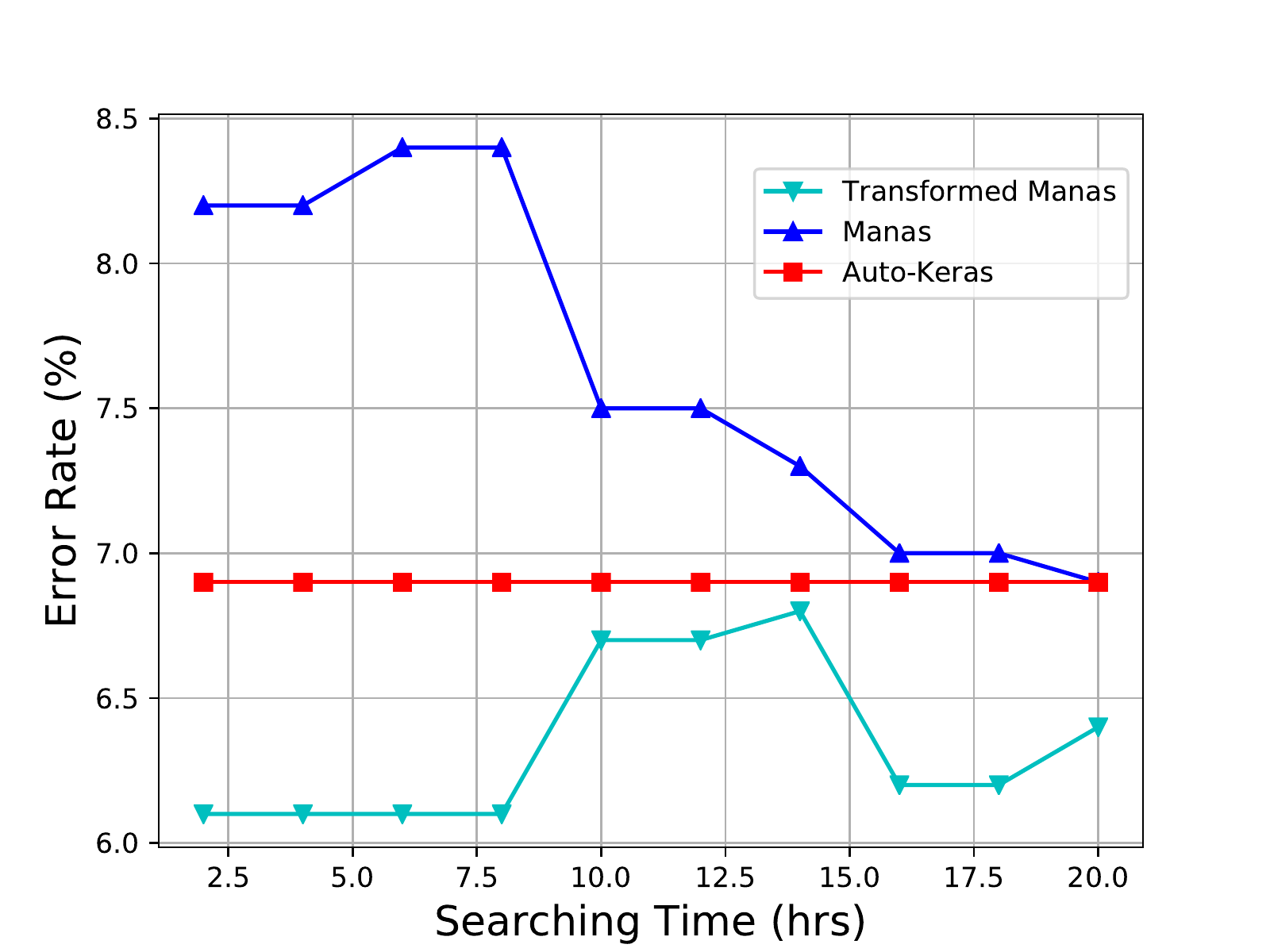}
	\caption{Breast Cancer Classification}
	\label{fig:breast}
\end{subfigure} 
\begin{subfigure}[b]{0.24\textwidth}
	\includegraphics[keepaspectratio = True, scale = 0.29]{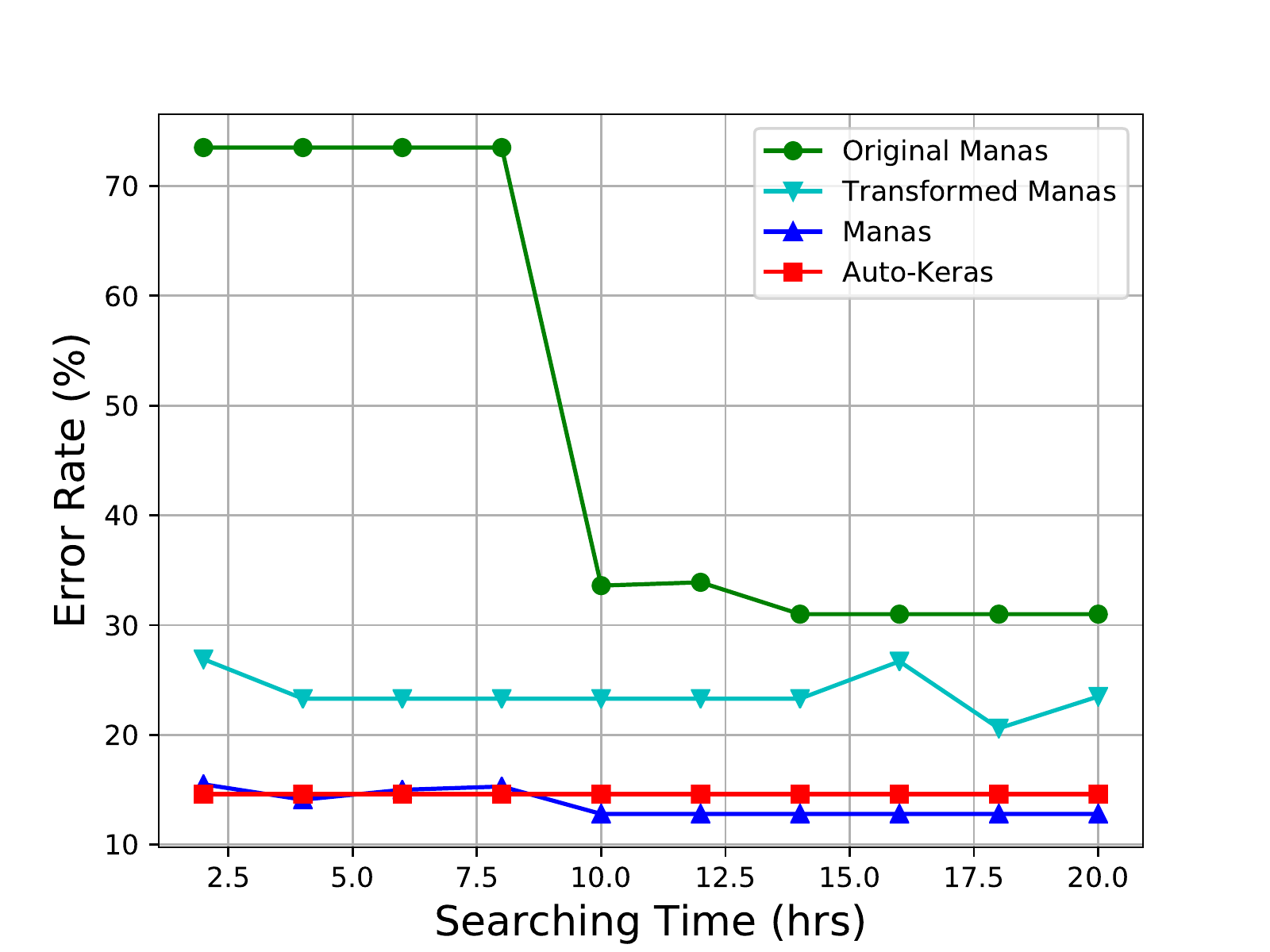}
	\caption{Flower Classification}
	\label{fig:flower}
\end{subfigure}
\begin{subfigure}[b]{0.24\textwidth}
	\includegraphics[keepaspectratio = True, scale = 0.29]{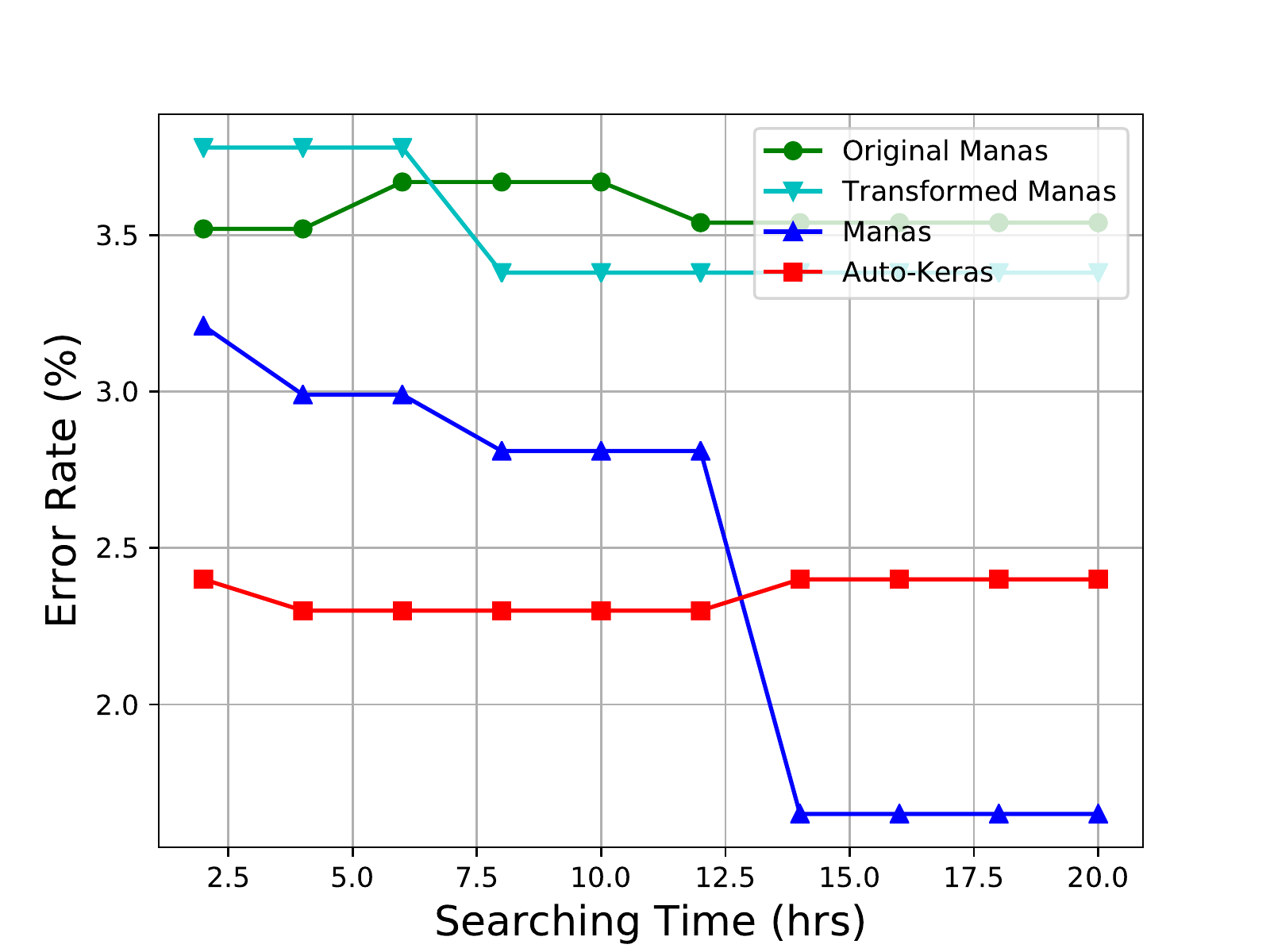}
	\caption{Malaria Classification}
	\label{fig:mala} 
\end{subfigure} 
\begin{subfigure}[b]{0.24\linewidth}
	\includegraphics[keepaspectratio = True, scale = 0.29]{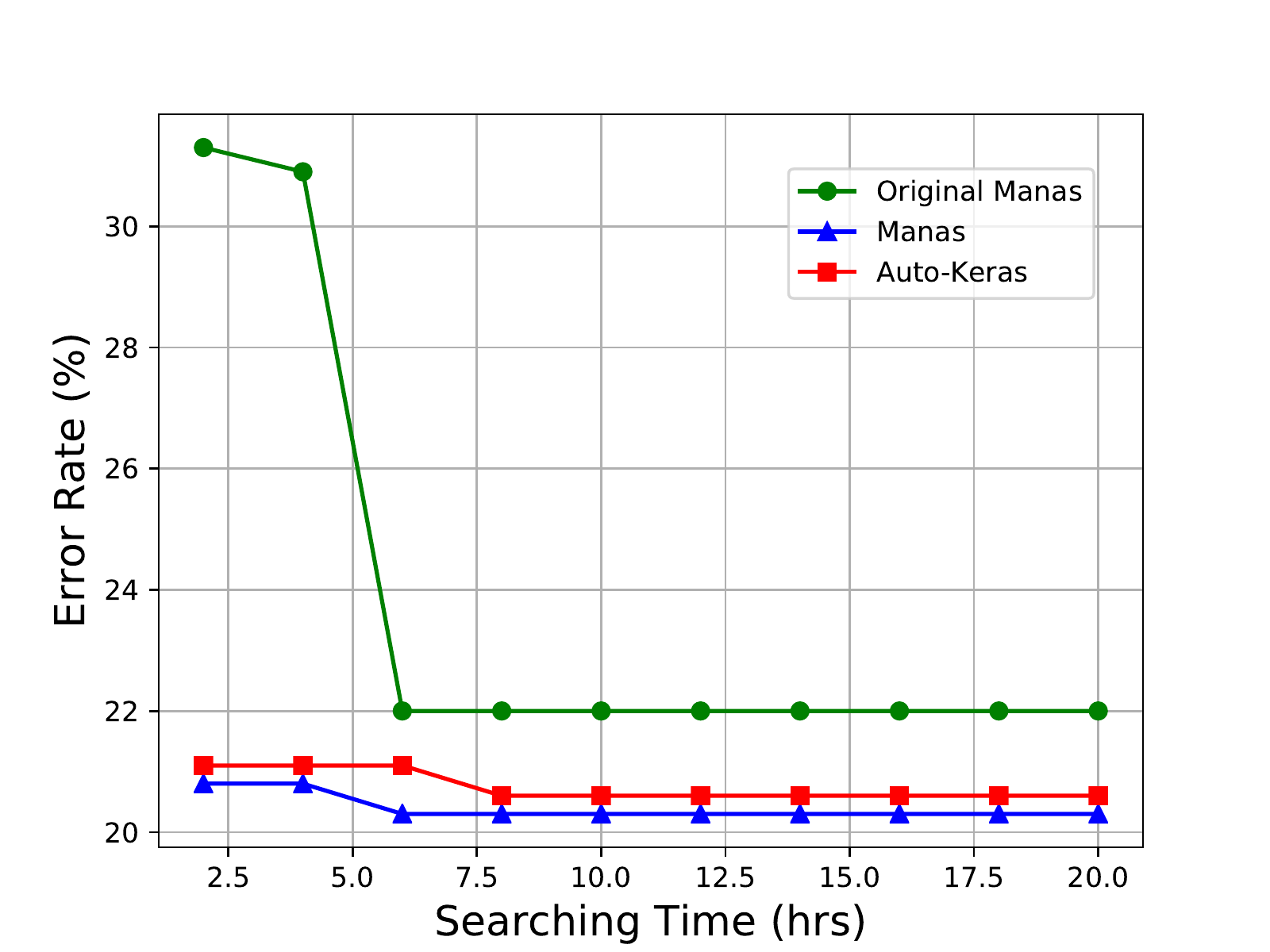}
	\caption{MNIST: Ham Classification}
	\label{fig:ham}
\end{subfigure} 
\begin{subfigure}[b]{0.24\linewidth}
	\includegraphics[keepaspectratio = True, scale = 0.29]{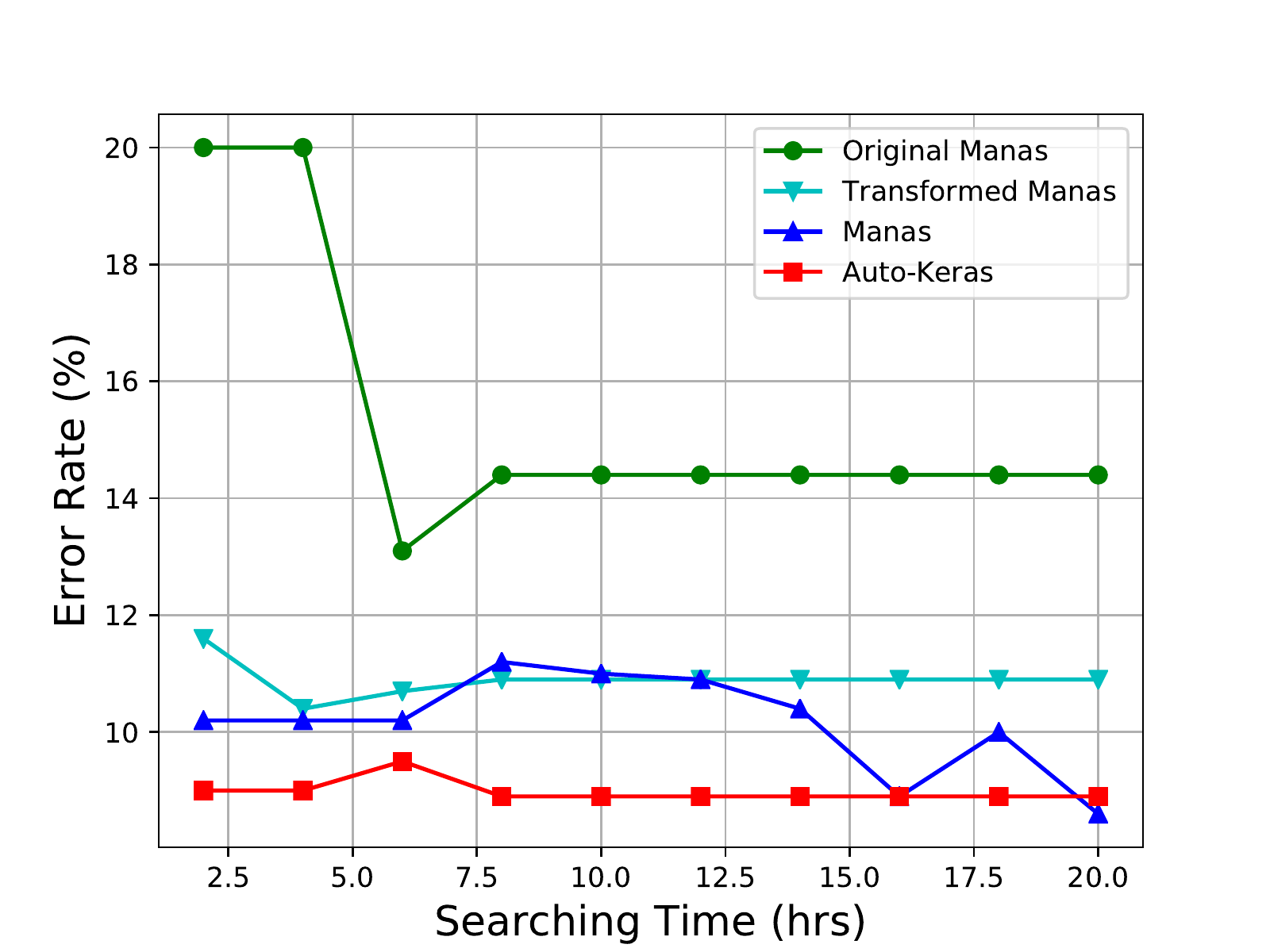}
	\caption{IIC Classification}
	\label{fig:iic}
\end{subfigure} 
\begin{subfigure}[b]{0.24\linewidth}
	\includegraphics[keepaspectratio = True, scale = 0.29]{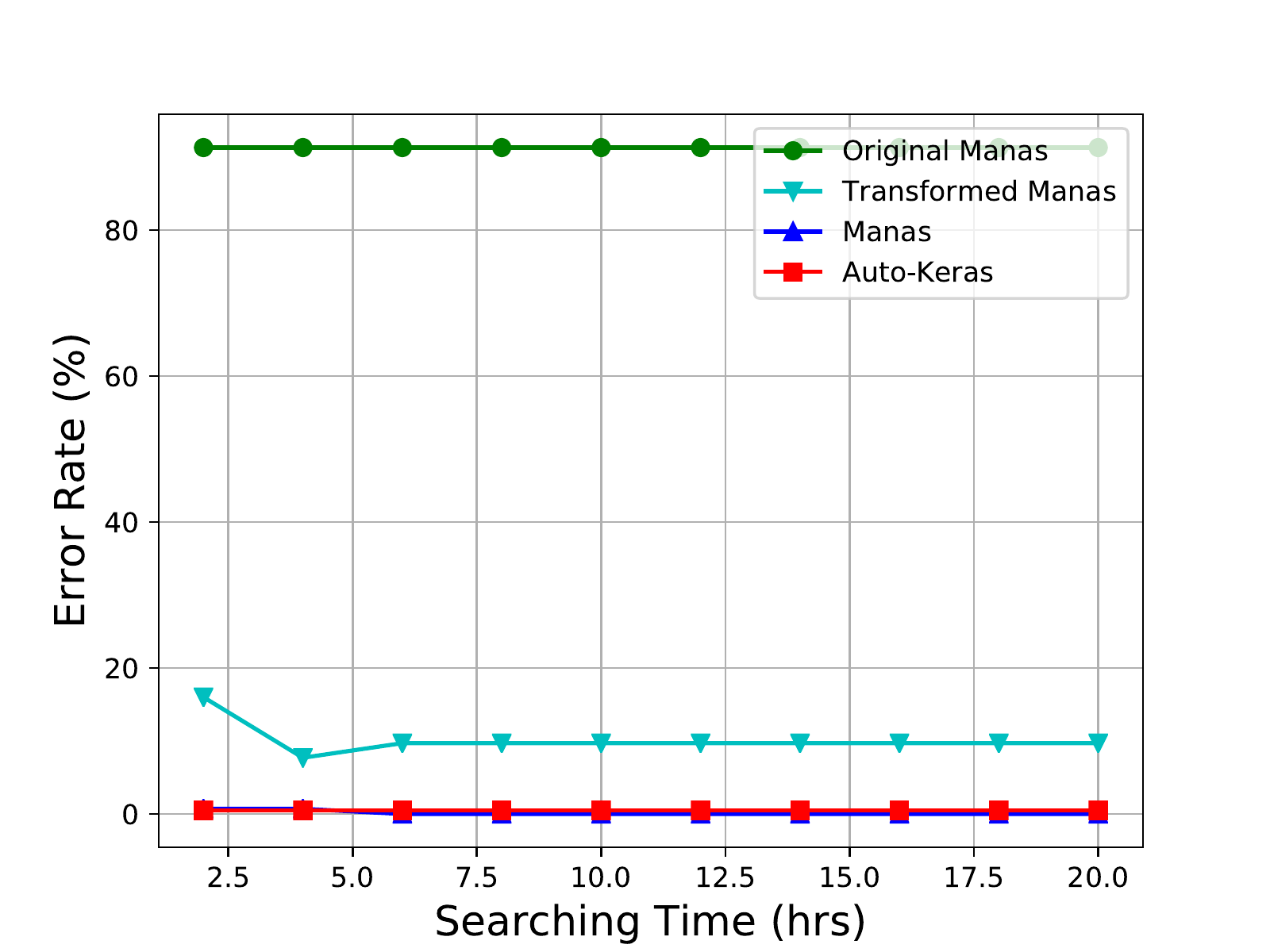}
	\caption{SD Classification}
	\label{fig:sd}
\end{subfigure} 
\begin{subfigure}[b]{0.24\linewidth}
	\includegraphics[keepaspectratio = True, scale = 0.29]{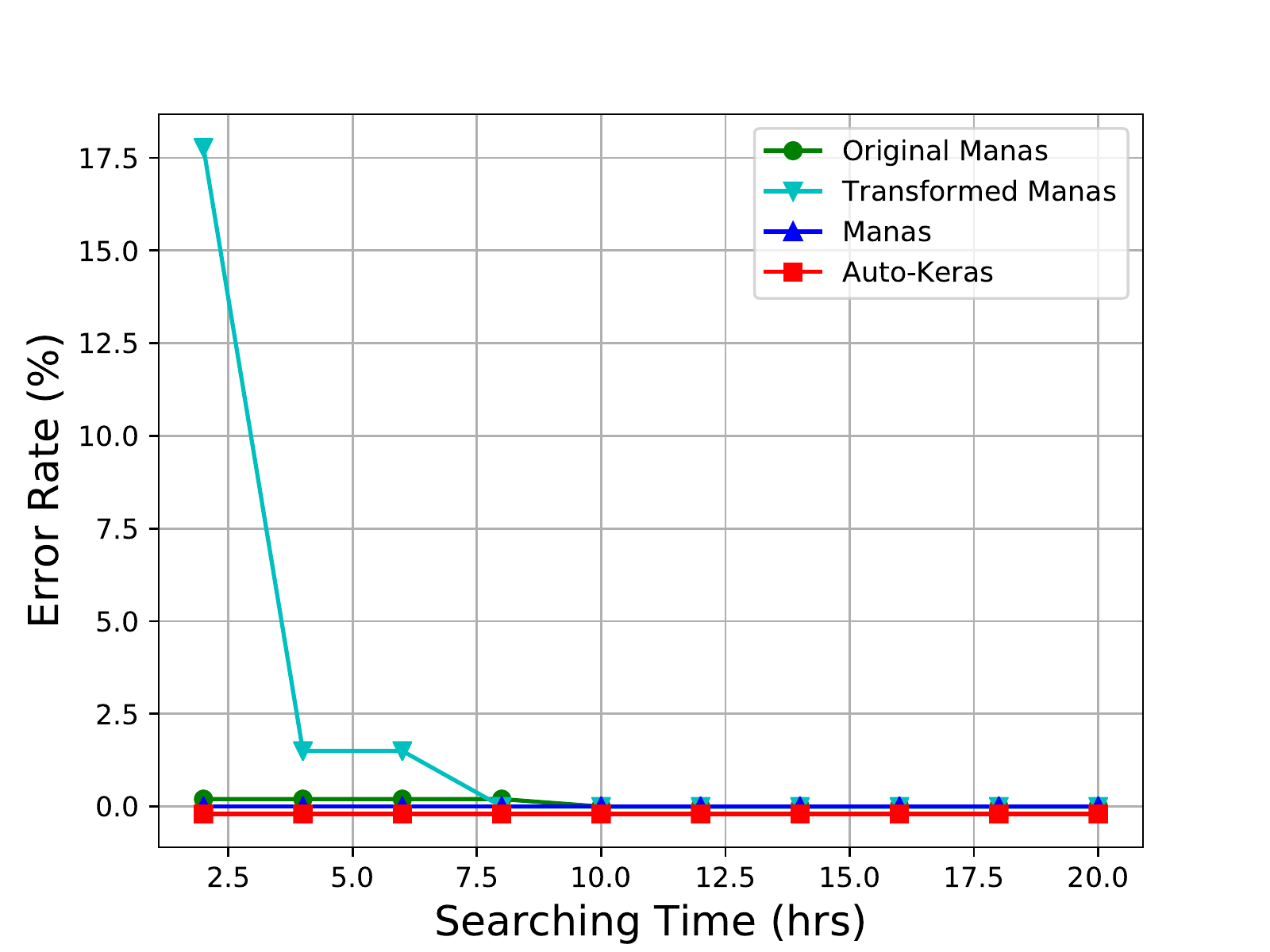}
	\caption{SL Classification}
	\label{fig:sl}
\end{subfigure} 
\begin{subfigure}[b]{0.24\linewidth}
	\includegraphics[keepaspectratio = True, scale = 0.29]{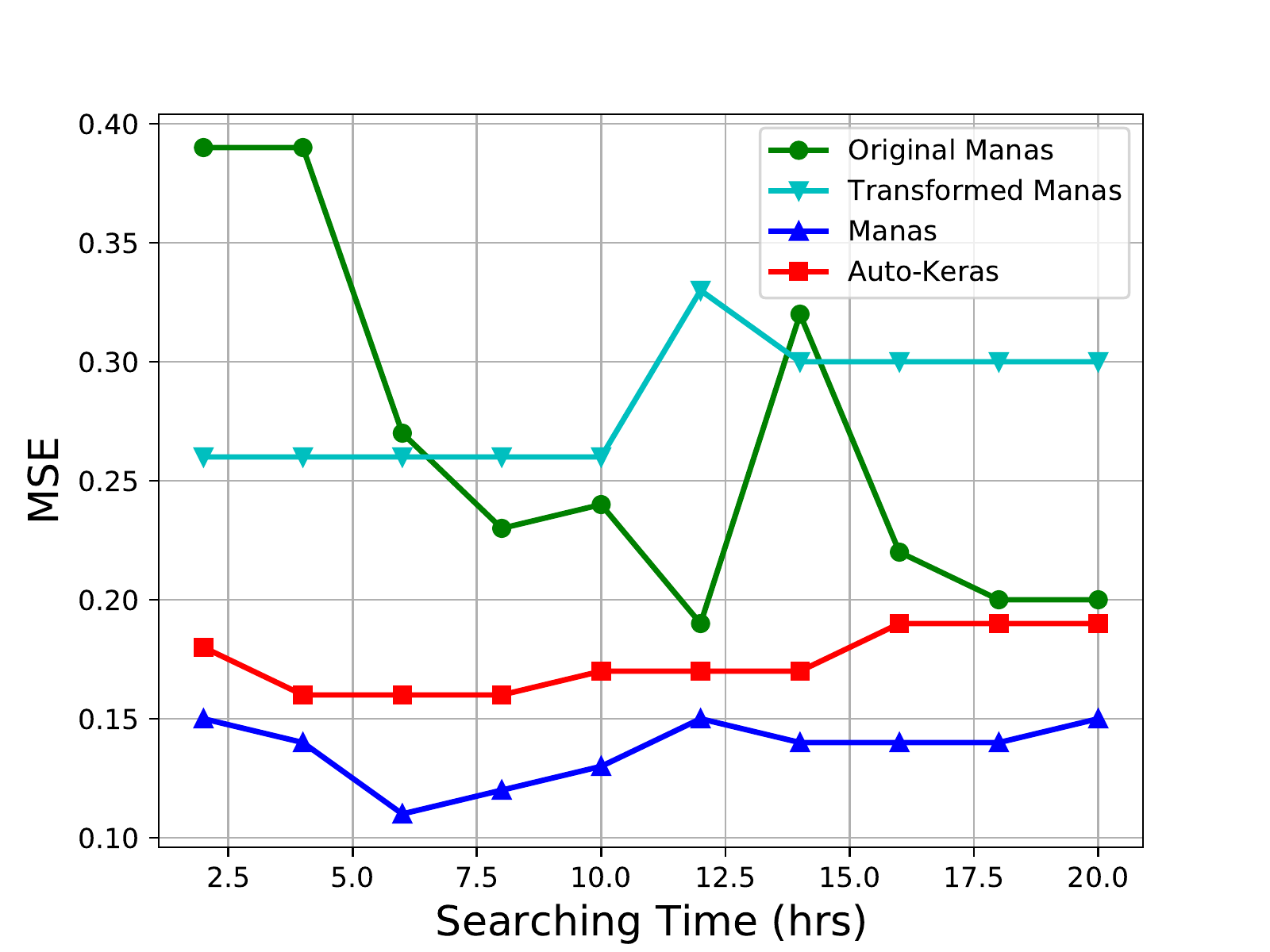}
	\caption{Blood Cell Regression}
	\label{fig:bcreg}
\end{subfigure} 
\begin{subfigure}[b]{0.24\textwidth}
	\includegraphics[keepaspectratio = True, scale = 0.29]{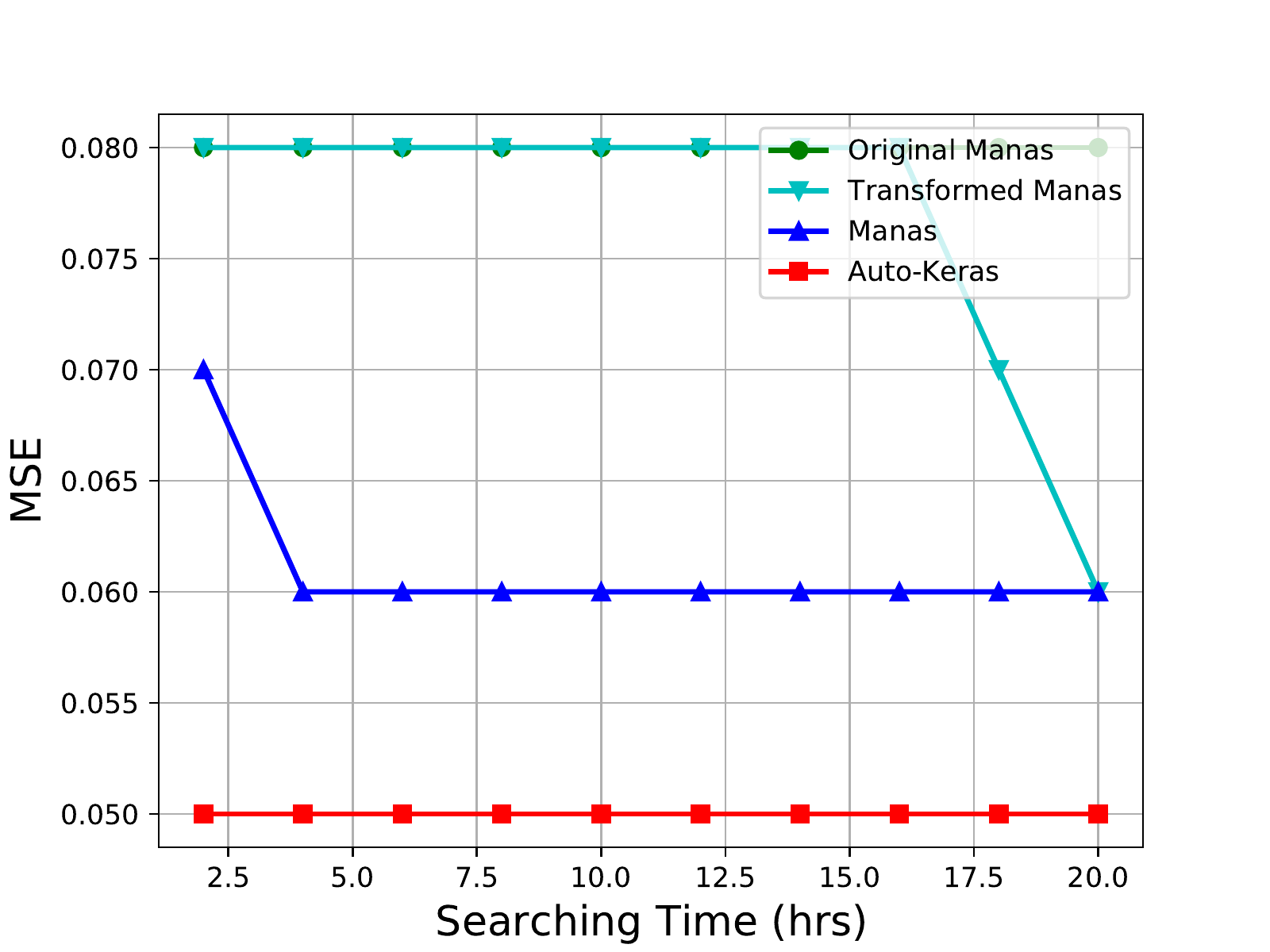}
	\caption{Breast Cancer Regression}
	\label{fig:breastreg}
\end{subfigure} 
\begin{subfigure}[b]{0.24\textwidth}
	\includegraphics[keepaspectratio = True, scale = 0.29]{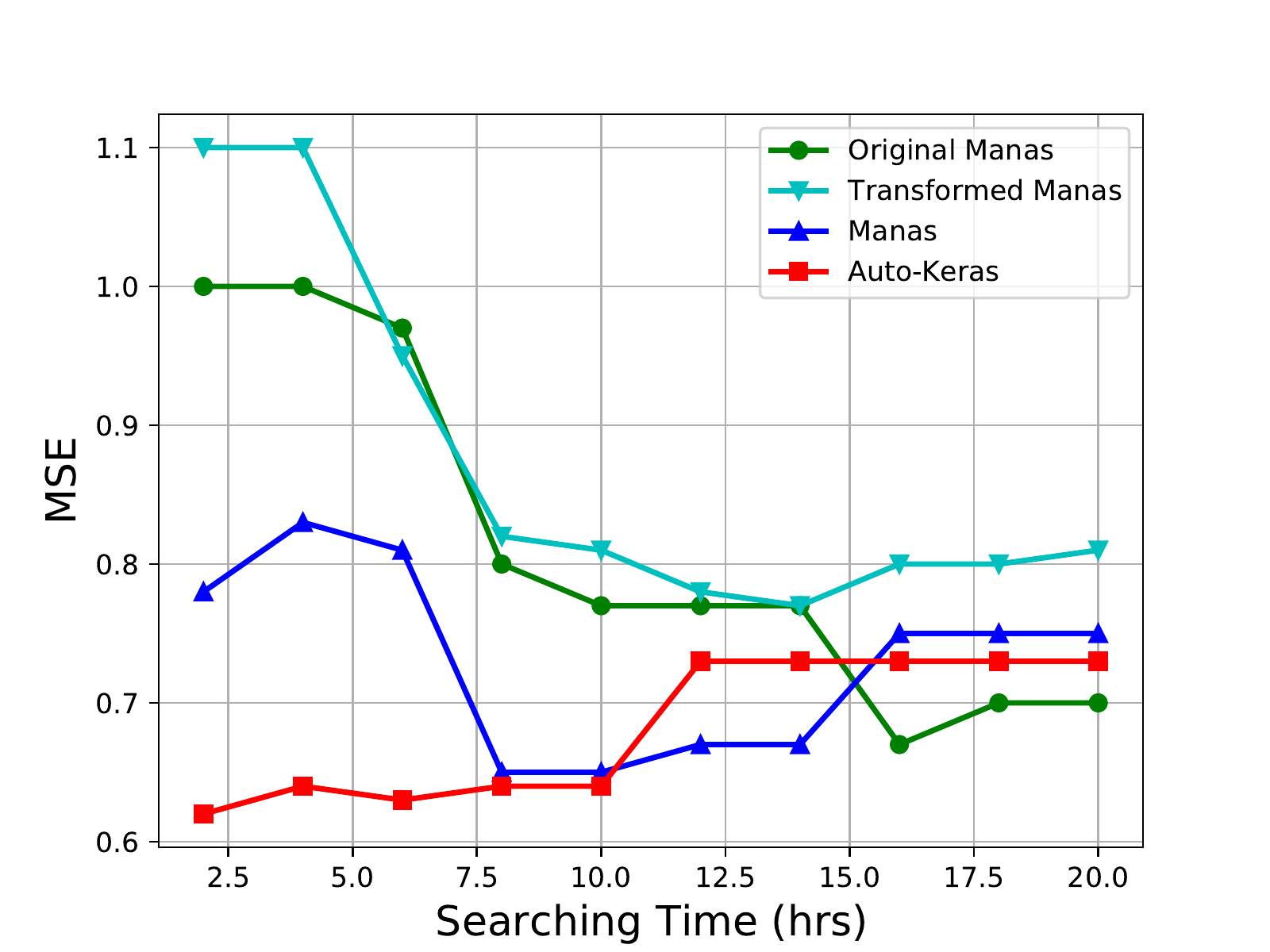}
	\caption{Flower Regression}
	\label{fig:flowerreg}
\end{subfigure} 
\begin{subfigure}[b]{0.24\textwidth}
	\includegraphics[keepaspectratio = True, scale = 0.29]{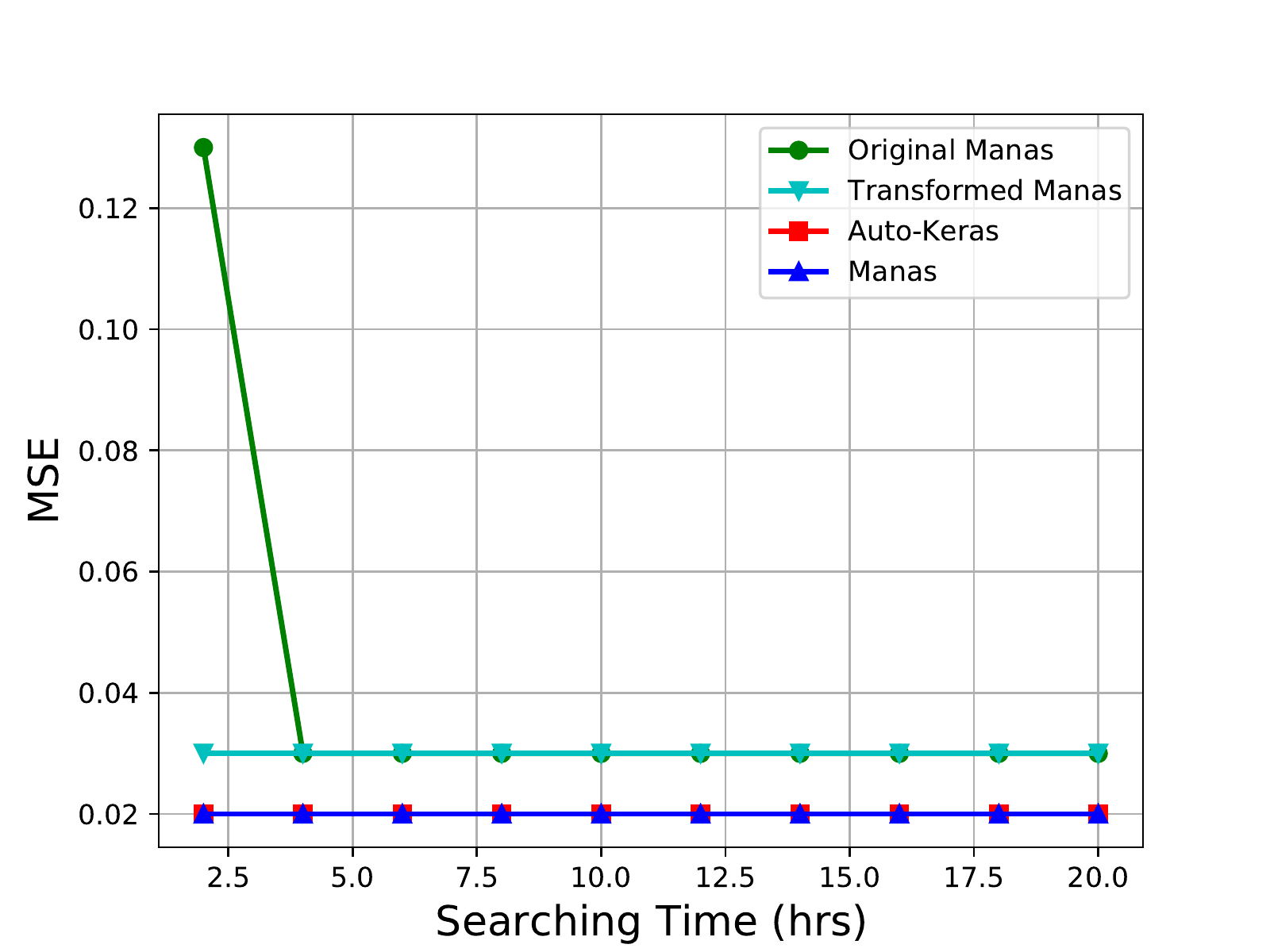}
	\caption{Malaria Regression}
	\label{fig:malareg}
\end{subfigure} 
\begin{subfigure}[b]{0.24\linewidth}
	\includegraphics[keepaspectratio = True, scale = 0.29]{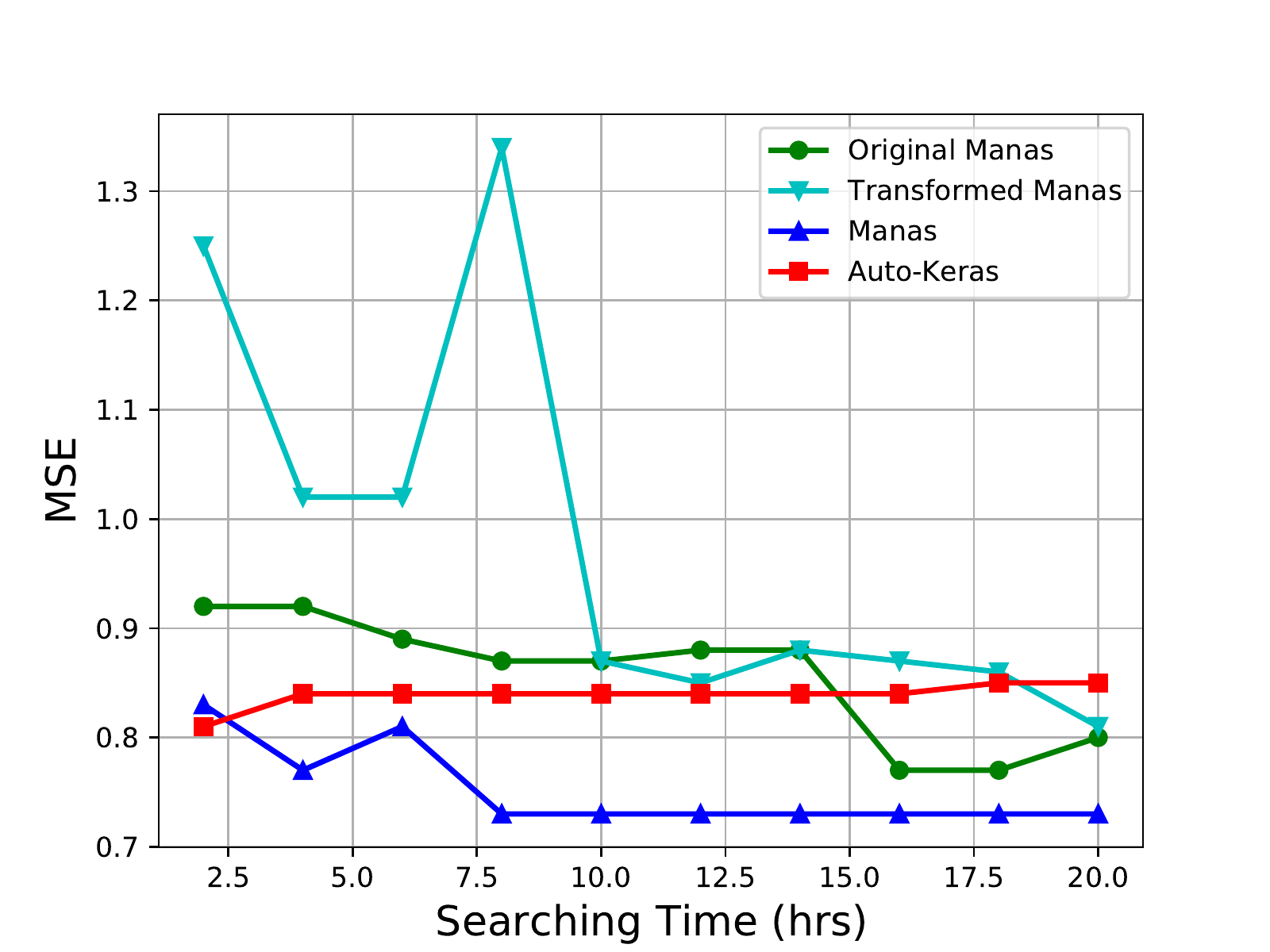}
	\caption{MNIST: Ham Regression}
	\label{fig:hamreg}
\end{subfigure} 
\begin{subfigure}[b]{0.24\linewidth}
	\includegraphics[keepaspectratio = True, scale = 0.29]{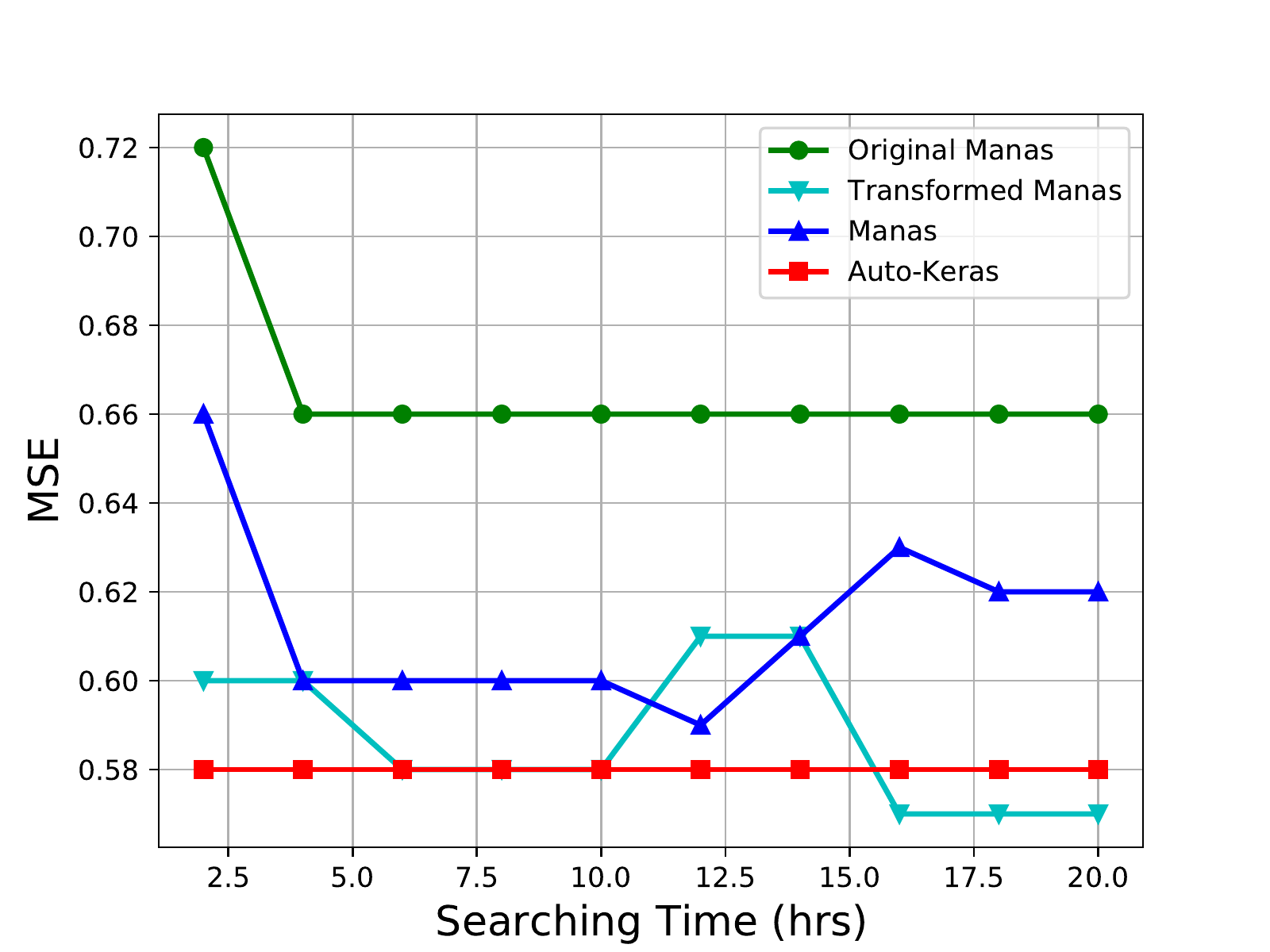}
	\caption{IIC Regression}
	\label{fig:iicreg}
\end{subfigure} 
\begin{subfigure}[b]{0.24\linewidth}
	\includegraphics[keepaspectratio = True, scale = 0.29]{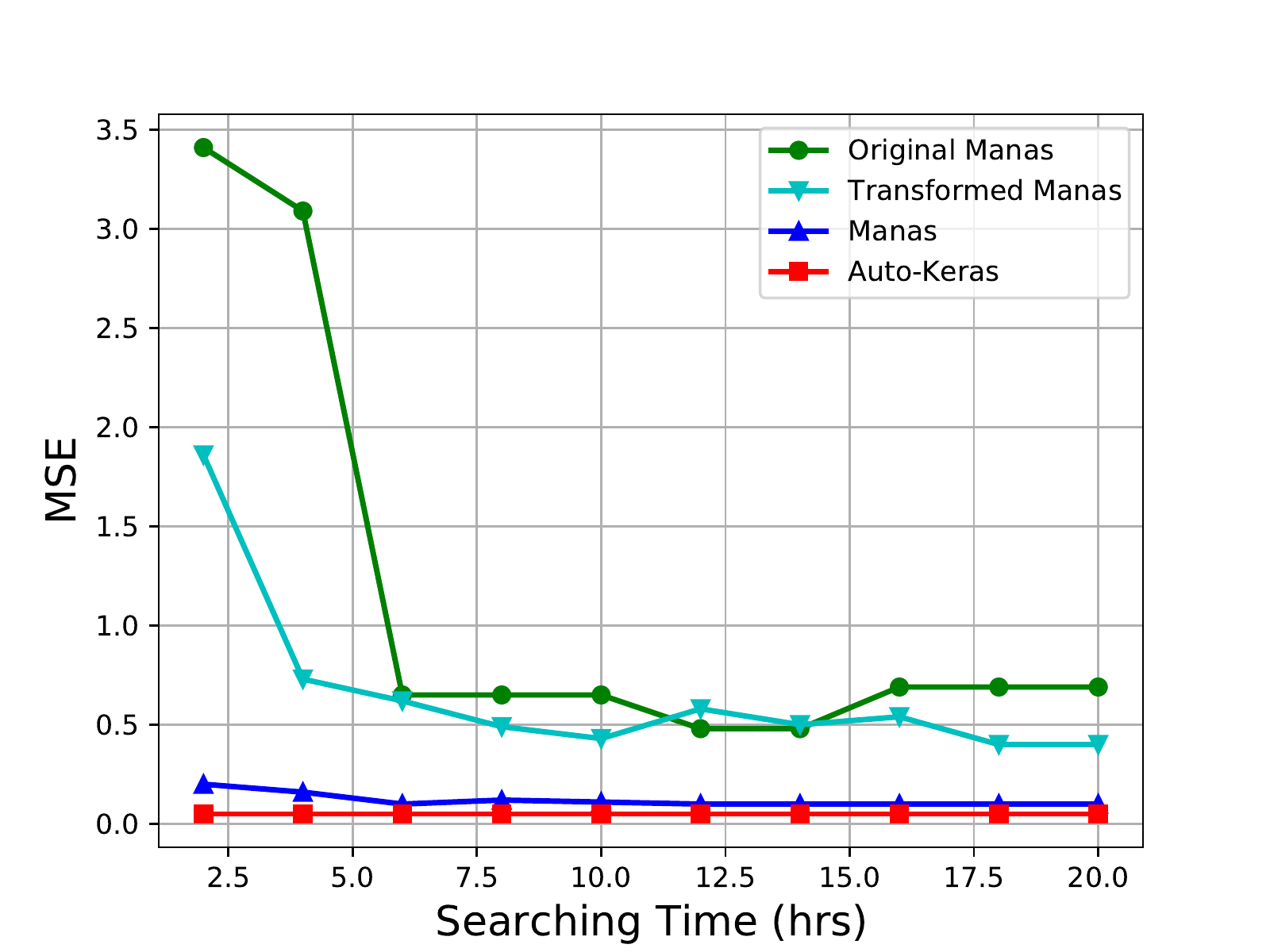}
	\caption{SD Regression}
	\label{fig:sdreg}
\end{subfigure} 
\begin{subfigure}[b]{0.24\linewidth}
	\includegraphics[keepaspectratio = True, scale = 0.29]{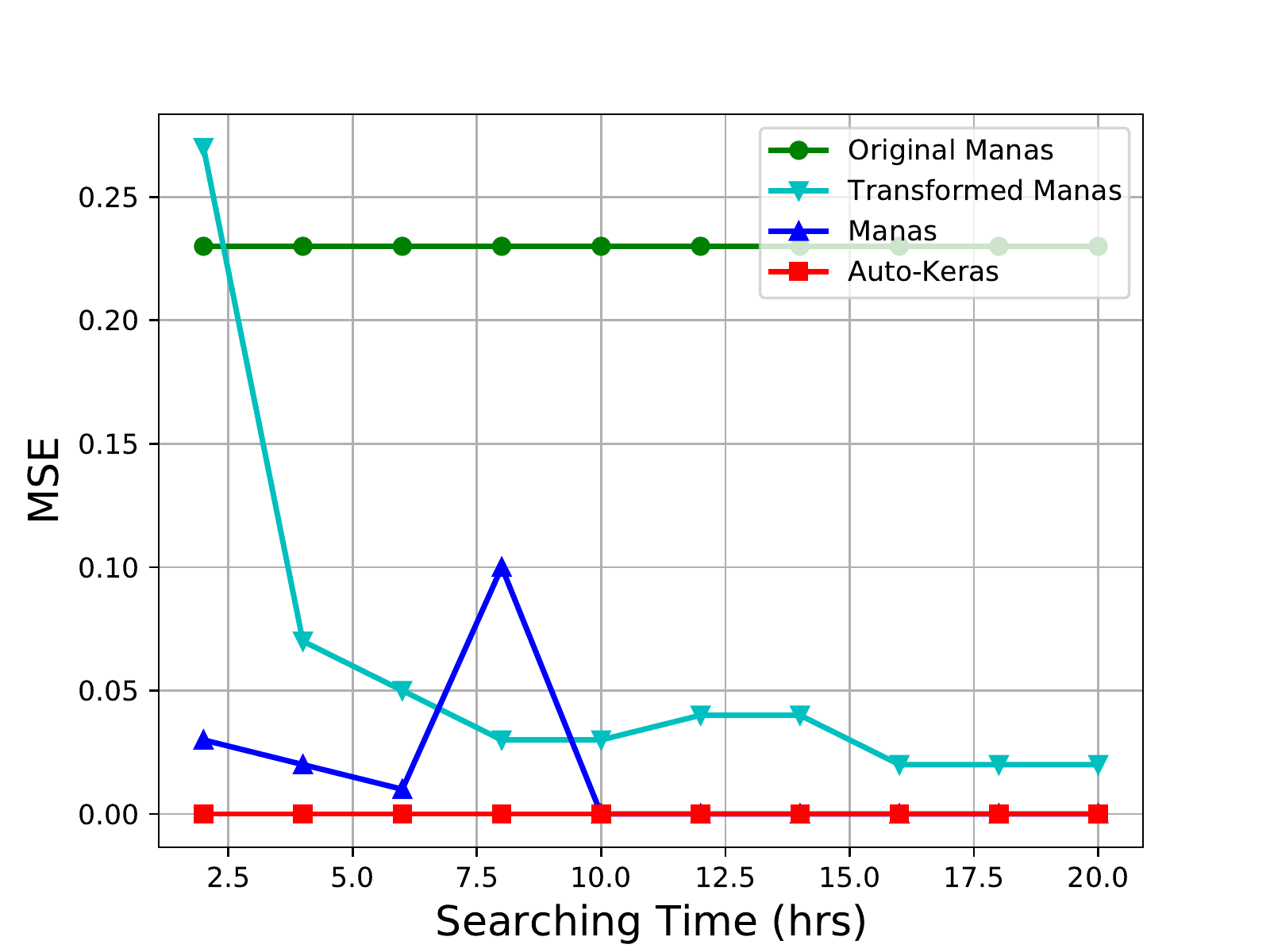}
	\caption{SL Regression}
	\label{fig:slreg}
\end{subfigure}
	\caption{Error Rate and MSE of \ak, \om, \tm, and \manas over time}
	\label{fig:results}
\end{figure*}

\subsection{Results}

\subsubsection{RQ1: How efficient is \manas?}

To evaluate the efficiency of \manas, we run both \manas and \ak on \nodata datasets for image classification and image regression.
We vary the search time from 2 hours to 20 hours, which are described in Figure~\ref{fig:results}. 
Table \ref{tbl:results} shows the error rate, MSE, the depth, the number of parameters, and the training speed 
of the best models of \manas and \ak.
By the best model, we mean one that has the lowest error rate or MSE after search timeout. We run both Manas and Auto-Keras 5 times for each dataset with random training and validation sets. In the table, we report the information of the lowest errors of Manas and Auto-Keras~\cite{cui2019fast}.
By comparing \manas and \ak, two conclusions can be drawn. 
 
First of all, Table \ref{tbl:results} shows that \manas produces models, which has lower error rate or MSE compared with \ak' models. The lower errors of \manas compared with \ak indicates that using mined models as the starting points for NAS can produce better models. Notably, \manas outperforms \ak by achieving \errorrate lower error rate on average. For some problems like \emph{IIC} and \emph{MNIST: HAM} for image classification task, the decrease of the error rate of \manas compared with \ak is small because the models created by \manas and \ak have reached the limit of error rate. Therefore, a small decrease in error rate is a major improvement. We use the error rate as the main evaluation metric to clearly point out this improvement of \manas to \ak. 
	
Secondly, \manas achieves lower errors with less complicated models compared to \ak. Using mined models on NAS significantly decreases the complexity of the produced models. On average, models generated by \manas are \depth less deep and \width less wide compared to the models generated by \ak for image classification task. Similarly, \manas creates models with \depthreg less deep and \widthreg less wide compare to \ak's models for image regression task. Simpler DNN models train faster and save more energy~\cite{iandola2016firecaffe} than the complex ones. Han \etal have shown that reducing the number of parameters of deep learning models can reduce the training time by 3$\times$ to 4$\times$, and energy comsumption by 3$\times$ to 7$\times$~\cite{han2015deep}. Table \ref{tbl:results} also shows that on average, \manas' models run faster than \ak' model \speed and \speedreg in image classification and image regression, respectively. 


\subsubsection{RQ2: How efficient are model transformation and optimizers modification?}

We create an ablation study to observe the efficiency of the model transformation and modifying optimizers via mining separately. 

\begin{itemize}
\item \om (OM) represents mined models + no transformation + no optimizer + NAS.
\item \tm (TM) represents mined models + transformations + no optimizer + NAS.
\item \manas (MN) represents for mined models + transformations + optimizers + NAS.
\item \ak (AK) represents NAS. 
\end{itemize}
We observe the error values of \om, \tm, \manas, and \ak by executing them on \nodata datasets for 20 hours for both image classification and image regression. To evaluate the efficiency of the model transformation, we compare \om and \tm. We compare \tm and \manas to evaluate the efficiency of optimizers modification. To evaluate the combination of all methods, we compare \manas and \ak.

Notice that we have applied the model transformation on \ak' models; however, this did not lead to any improvement because those default models use BN layers or Dropout layers appropriately. Figure \ref{fig:breast} does not have the \om series because the best models of \om and \tm are the same. In other words, the model already contains all the transformation constraints, so there is no transformation applied to the model. Similarly, Figure \ref{fig:ham} does not have the \tm series because the optimizers of the best model of \tm and \manas are the same. The selected model's optimizer of this problem is not available; therefore, we use the default optimizer of \ak for this model. From the results are shown in Figure \ref{fig:results}, we draw four observations.


First of all, model transformation increases the performance of \manas in terms of errors and converge time. Figures \ref{fig:bc}, \ref{fig:flower}, \ref{fig:iic}, \ref{fig:sd} show that \tm achieve lower errors than \om most of the time. The activation layers, the GAP layer, and the dropout layers contribute to the better errors of \tm compared to \om. For example, the dropout layers can prevent overfitting, which decreases the errors. The model transformation also helps \manas to converge faster. As we can observe Figures \ref{fig:bc}, \ref{fig:flower},  \ref{fig:iic}, \ref{fig:sd}, \om has higher errors compared to \tm in the first 2 hours of training, which indicate the \tm converge faster than \om. \tm has a fast converge speed thanks to BN.

%

Secondly, mined optimizers help \manas to reduce errors and converge faster. 
In most of problems, the best models of \manas have lower errors than the best models of \tm. Moreover, \manas can converge easier with the mined optimizers. 
For instance, the Figures \ref{fig:mala}, \ref{fig:iic}, \ref{fig:sd} shows that \tm has trouble converging. 
After 8 hours of searching, \tm cannot find out better models while \manas succeeds. \tm has trouble in converging since it does not have an suitable optimizer for those problems like \manas. 

Thirdly, \ak may have lower errors than \manas in the first few hours; however, in the last hours, \manas often finds out better models than \ak. For example, Figures \ref{fig:breast}, \ref{fig:flower}, \ref{fig:mala}, and \ref{fig:hamreg} show that \ak has the errors at the beginning; however, as time gone we have not seen any improvement in \ak results. The reason for this problem is that \ak starts with very complicated models that take a lot of time to train; therefore, the number of models is searched by \ak are small, which decreases the chance to find out good models of \ak. \manas starts with simple models, which trains faster. Therefore, \manas may not be off to a good start, but in the end, it still outperforms \ak. The observation shows the benefit of using simple mined models in NAS.

Lastly, Figure \ref{fig:results} indicates that \manas obtains lower error rates than \ak in almost different time periods. As we can observe in Figures \ref{fig:bc} and \ref{fig:bcreg}, \manas always outperforms \ak in terms of error rate in the 20-hours of searching.  Figure \ref{fig:results} shows that the longer searching time sometimes gives worse models. During the searching process, the NAS estimates the errors of searched models and selects the best one. However, the estimation may not be accurate, leading to an incorrect choice. This problem of NAS indicates the importance of using a simple model as a starting point. Simple models can train faster that increases the chance for NAS to search for more models. Since the estimation of NAS can be incorrect; thus, if we increase the number of searched model, we can increase the chance to obtain better models. For example, after 8-hours of searching, NAS goes wrong with \emph{IIC} problem when it produces worse models than before. However, when we keep searching for new models, NAS gradually fixes the problem to obtains a good model. 

\begin{table}
	\centering
	\caption{Efficiency of Model Matching}
	\setlength{\tabcolsep}{1.7pt}
\begin{tabular}{|c|r|r|r|r|r|r|r|r|r|r|r|r|r|r|r|r|}
	\hline
	Data  & \multicolumn{2}{c|}{\begin{tabular}[c]{@{}c@{}}Blood\\ Cell\end{tabular}} & \multicolumn{2}{c|}{\begin{tabular}[c]{@{}c@{}}Breast\\ Cancer\end{tabular}} & \multicolumn{2}{c|}{Flower} & \multicolumn{2}{c|}{IIC} & \multicolumn{2}{c|}{\begin{tabular}[c]{@{}c@{}}Mala-\\ ria\end{tabular}} & \multicolumn{2}{c|}{Ham} & \multicolumn{2}{c|}{SD} & \multicolumn{2}{c|}{SL} \\ \hline
	Task  & IC                                  & IM                                  & IC                                    & IM                                   & IC           & IM           & IC          & IM         & IC                                  & IM                                 & IC          & IM         & IC         & IM         & IC         & IM         \\ \hline
	Total & \multicolumn{16}{c|}{793}                                                                                                                                                                                                                                                                                                                                                   \\ \hline
	MC    & 5                                   & 5                                   & 5                                     & 7                                    & 18           & 6            & 11          & 46         & 5                                   & 7                                  & 6           & 11         & 69         & 4          & 5          & 5          \\ \hline
	MF    & -                                   & -                                   & -                                     & -                                    & -            & -            & -           & 12         & -                                   & -                                  & -           & -          & 12         & -          & -          & -          \\ \hline
\end{tabular}
\label{tbl:matchingtab}
\centering

\scriptsize \* In each cell, IC, IM, MM, and MF represent image classification, image regression, model clustering and model filtering, respectively. The unit of all the data in the table is the number of models.
\end{table}

\subsubsection{RQ3: How efficient is model matching?} 
\begin{table}
\centering
\caption{\manas vs \ak for Well-known Problems}
\setlength{\tabcolsep}{1.2pt}
\footnotesize
\begin{tabular}{|c|r|r|r|r|r|r|r|r|}
\hline
\multirow{2}{*}{Data}  
                      & \multicolumn{2}{c|}{\begin{tabular}[c]{@{}c@{}}Error Rate\\ (\%)\end{tabular}} & \multicolumn{2}{c|}{\begin{tabular}[c]{@{}c@{}}Depth\\ (layers)\end{tabular}} & \multicolumn{2}{c|}{\begin{tabular}[c]{@{}c@{}}Param \#\\ (million)\end{tabular}} & \multicolumn{2}{c|}{\begin{tabular}[c]{@{}c@{}}Speed\\ (epoch/min)\end{tabular}} \\ \cline{2-9} 
                      & AK                                     & MN                                    & AK                                    & MN                                    & AK                                       & MN                                     & AK                                      & MN                                     \\ \hline
CIFAR10               & 7                                      & 7.7 ($\uparrow${
 10.0\%})                                   & 21                                    & 10 ($\downarrow${
 52.4\%})                                   & 19.4                                     & 0.8 ($\downarrow${
 95.9\%})                                   & 1                                       & 2.9 ($\uparrow${
 190.0\%})                                   \\ \hline
Fashion               & 5.2                                    & 11($\uparrow${
 111.5\%})                                    & 22                                    & 9 ($\downarrow${
 59.1\%})                                     & 14.7                                     & 1.8 ($\downarrow${
 87.8\%})                                   & 2.1                                     & 1.8 ($\downarrow${
 14.3\%})                                    \\ \hline
MNIST                 & 0.5                                    & 0.7 ($\uparrow${
 40.0\%})                                  & 23                                    & 6 ($\downarrow${
 73.9\%})                                    & 19.3                                     & 2.9 ($\downarrow${
 85.0\%})                                   & 1.3                                     & 3.1 ($\uparrow${
 138.5\%})                                   \\ \hline \hline

Avg                 & 4.2                                    & 6.5 ($\uparrow${
 54.8\%})                                   & 22                                    & 8.3 ($\downarrow${
 62.3\%})                                    & 17.8                                     & 1.8  ($\downarrow${
 89.9\%})                                  & 1.5                                     & 2.6 ($\uparrow${
 73.3\%})                                   \\ \hline
\end{tabular}
\label{tbl:wkresults}
\centering

\scriptsize \* In each cell, Avg, AK and MN represent average, \ak, and \manas, respectively. 
\end{table}


The goal of model matching is not only selecting good default models for \manas but also reducing the number of default models. From Table \ref{tbl:results}, we observe the efficiency of model matching, when \manas can outperform \ak in many different perspectives. Table \ref{tbl:matchingtab} shows that model matching, including model clustering and model filtering, can significantly reduce the number of default models for \manas.To balance the time that \manas spends between initial architectures and NAS, we applied the models filtering when the number of initial neural networks is larger than 40 since \manas often takes more than 10 hours, half of the amount of time used in our evaluations to complete training these neural networks. We use all 793 mined models as input for each problem, which may take few days to complete training. By using model matching, we decrease the number of default models by 99\% on average. Taking \textit{SD} in image classification as an example, after using DNN clustering, there are remaining 69 default models. 
Therefore, we use DNN filtering on this dataset to reduce the number of models of 
\textit{SD} from 69 models to 12 models, which eliminates 82.6\% number of the DNN model.

\subsubsection{RQ4: How efficient is \manas for well-known problems?}
\label{sec:well-known}

We also evaluate \manas with the well-known datasets like FASHION, MNIST, or CIFAR10 for the image classification task. Table \ref{tbl:wkresults} shows the error rate, depth, number of parameters, and speed of the best models of \manas and \ak. As can be observed, \ak achieve better error rates on these datasets than \manas; however, \ak produces larger models to achieve these error rates while \manas uses much smaller models to get close to the error rates of \ak in CIFAR10 and MNIST problems. Particularly, \manas' models are 62.3\% shorter and 89.9\% than \ak’ models on average, which increases the training speed of \manas' models by 73.3\% compared to \ak’ models. \ak achieves better error rates than \manas since \ak uses ResNet and DenseNet as initial models. These neural networks are well-tuned to achieve outstanding results on these datasets, which once again shows that using good initial architecture optimizes NAS.

	\section{Limitations and Threats To Validity}
\label{sec:limit}
\subsection{Limitations} In this work, \manas directly derives the data characteristics from the image dataset, which is limited to the image classification and image regression problems. 
We believe that \manas is not directly applied to other problems such as natural language processing (NLP) or video classification; however, our approach of mining models to identify a good starting point candidate should be applicable to any AutoML problems. For example, data characteristics of NLP problems also include the input shape, which are input time steps and the number of features, and the total number of output classes. Time steps represent the maximum length of the input sequence, which could either be the number of words or the number of characters depending on what we want. The number of features is the number of dimensions we feed at each time step. We can analyze the input layer and output layer to extract data characteristics from models. We can analyze the input dataset to obtain its data characteristics. The extracted data characteristics can be used to find better starting points for NLP problems in AutoML systems.

\manas can only work with few kinds of layers since it only uses the layers that
\ak supports. 
This limitation can decrease the performance of \manas because if \ak supported 
more layers, we can mine more kinds of models from the software repositories. 

\subsection{Threats to Validity}
\subsubsection{Internal validity} We have tried our best to obtain the results of \manas and \ak on as many datasets as possible. Because of the time limit, \manas is currently evaluated on \nodata datasets. All the source code, trained models, datasets, and evaluation data are public for reproduction to mitigate these threats.


\subsubsection{External validity}  First of all, \manas only focuses on image classification/regression problems. We rely on meta-features to find good starting points for NAS; therefore, one possible threat is that meta-features (data characteristics) do not work for other types of problems. However, Auto-Sklearn~\cite{feurer2015efficient} and Auto-Sklearn 2.0~\cite{feurer2020auto} mitigate this threat by showing that using meta-features can increase the performance of AutoML systems in terms of training speed and accuracy on structured datasets. Secondly, \manas only focuses on CNN. Thus, another threat is that the model transformation approach does not work for other types of models. Nevertheless, Cambronero \etal~\cite{cambronero2020ams} propose AMS showing that using unspecified complementary and functionally related API components can improve the performance of AutoML systems for classical models such as Linear Regression or Random Forest. The difference between \manas and AMS is that AMS applies these transformations to search space while \manas applies these transformations to default models.

	\section{Related Work}
\label{sec:related}
\subsection{Neural Architecture Search} NAS is a technique for automatically finding appropriate neural architectures which can outperform most of the hand-designed neural networks. Specifically, NAS needs a the training dataset as the input to create a powerful neural architecture. There are many different approaches for a NAS system; however, most of them have three main components which are search space, search strategy, and optimization strategy. The search space represents the search boundary of a NAS system limiting what kinds of neural network can be searched and optimized. For instance, Baker \etal~\cite{DBLP:conf/iclr/BakerGNR17} use the convolutional architecture with pooling, linear transformations as a search space. Around the same time, Zoph \etal~\cite{DBLP:conf/iclr/ZophL17} use a similar search space; however, the authors use more skip connection for the search space. The search strategy is used to search appropriate models in a defined search space. There are many approaches to search models such as reinforcement learning~\cite{DBLP:conf/iclr/BakerGNR17, cai2018efficient, zhong2018practical, DBLP:conf/iclr/ZophL17, zoph2018learning} or evolutionary algorithms~\cite{real2017large, suganuma2017genetic, xie2017genetic}. This optimization strategy supports NAS to guide the network search process. The optimization strategy evaluates a searched model with training data without training these models. Recently, many methods are proposed to optimize NAS~\cite{kandasamy2018neural, nayman2019nips, peng2019nips}. In our work, \manas mines the neural networks from repositories to enhance the power of NAS by supporting it to have a better starting point. 

\subsection{AutoML} AutoML is a process for constructing an appropriate model architecture for a specific problem automatically. Many features that AutoML can provide to deep learning users, such as automated data augmentation, automated hyperparameter tuning, or automated model selection. A lot of AutoML systems have been created like Auto-WEKA~\cite{thornton2013auto} on top of WEKA~\cite{hall2009weka, witten2016data}, Auto-Sklearn~\cite{feurer2015efficient} on top of Scikit-learn, which support deep learning user to automate tuning hyperparameter and model selection. Some other AutoML systems can support deep learning users to automate optimizing the full ML pipeline. For instance, TPOT~\cite{olson2016evaluation, olson2016automating} uses evolutionary programming to optimize ML software. However, a main disadvantage of these systems are very slow because of the high GPU computation requirement. Recently, \ak is created to handle this problem, which has implemented network morphism~\cite{DBLP:journals/corr/ChenGS15, wei2016network} to reduce the searching time of NAS. Network morphism is a technique to morph a neural architecture without changing its functionality. Nevertheless, even though \ak apply network morphism technique, it still takes a lot of GPU computation. Our approach uses DNN model mining and common layer patterns to enhance the performance of AutoML system.

\subsection{Mining Software Repositories} Cambronero \etal \cite{cambronero2019autogenerating} proposed AL, a system that leverages existing machine learning code from repositories to synthesize final pipelines. AL can generate ML pipelines for a wide range of problems without any manual selection. Cambronero \etal\cite{cambronero2020ams} also proposed AMS, which automated generates new search space for AutoML systems by utilizing source code repositories. The new search space is created based on an input ML pipeline, which increases the performance of AutoML systems. However, these only operate classical machine learning models, whereas \manas works with neural networks.

\subsection{Meta-features} Auto-Sklearn~\cite{feurer2015efficient} uses 38 meta-features of structured datasets to find a better starting point. Feurer \etal~\cite{feurer2020auto} proposes Auto-Sklearn 2.0, which reduces the number of meta-features to three, including the number of data points, the number of
features, and the number of classes. The reasons for the reduction are that good meta-features are time-consuming and memory-consuming to generate. We also do not know which meta-features work best for which problem. Unlike Auto-Sklearn and Auto-Sklearn 2.0, \manas uses meta-features to find a better starting point for NAS, which works for neural networks. Moreover, \manas also proposes the meta-features, which helps NAS find a better starting point for image datasets.

	\section{Conclusion}
\label{sec:conclude}
We present \manas, a technique for NAS, which uses the mining technique to assist NAS. 
The key idea of \manas is to mine models from repositories to enhance NAS. 
In particular, we use CNN models mined from software repositories as the default 
model of NAS. 
From a large number of models, we use the model matching approach to find good models for a problem. 
We also apply some transformations for those models to enhance their performances. 
With better default models, \manas can increase NAS's performance, 
which leads to better CNN models as search results. 
Our experiment shows that \manas can produce better CNN models in terms of the error rate and MSE, 
the model complexity, and the training speed than \ak.  
Future work will involve extending \manas to problems beyond those tackled in this paper, 
such as video classification. We can utilize the code change patterns~\cite{DilharaICSE2022RepetitiveCode} in ML programs to improve the results. Moreover, the proposed approach of \manas can be applied to other automated tools, e.g., AutoAugment~\cite{cubuk2019autoaugment} of other components in ML pipeline~\cite{biswas22art}. The technique can also be utilized to use AutoML to address the other problems in ML, such as fairness bug~\cite{biswas20machine}.


\begin{acks}
This work was supported in part by US NSF grants CNS-21-20448, CCF-19-34884, and Facebook Probability and Programming Award (809725759507969).  We also thank the reviewers for their insightful feedback for improving the paper.
\end{acks}
	\balance
	
	\bibliographystyle{ACM-Reference-Format}
	\bibliography{refs}


\begin{thebibliography}{62}


\ifx \showCODEN    \undefined \def \showCODEN     #1{\unskip}     \fi
\ifx \showDOI      \undefined \def \showDOI       #1{#1}\fi
\ifx \showISBNx    \undefined \def \showISBNx     #1{\unskip}     \fi
\ifx \showISBNxiii \undefined \def \showISBNxiii  #1{\unskip}     \fi
\ifx \showISSN     \undefined \def \showISSN      #1{\unskip}     \fi
\ifx \showLCCN     \undefined \def \showLCCN      #1{\unskip}     \fi
\ifx \shownote     \undefined \def \shownote      #1{#1}          \fi
\ifx \showarticletitle \undefined \def \showarticletitle #1{#1}   \fi
\ifx \showURL      \undefined \def \showURL       {\relax}        \fi
\providecommand\bibfield[2]{#2}
\providecommand\bibinfo[2]{#2}
\providecommand\natexlab[1]{#1}
\providecommand\showeprint[2][]{arXiv:#2}

\bibitem[\protect\citeauthoryear{{Anonymized}}{{Anonymized}}{2015}]%
        {so49226447}
\bibfield{author}{\bibinfo{person}{{Anonymized}}.}
  \bibinfo{year}{2015}\natexlab{}.
\newblock \bibinfo{booktitle}{\emph{Resnet network doesn't work as expected}}.
\newblock
\urldef\tempurl%
\url{https://stackoverflow.com/questions/49226447/resnet-network-doesnt-work-as-expected}
\showURL{%
\tempurl}


\bibitem[\protect\citeauthoryear{{Anonymized}}{{Anonymized}}{2021}]%
        {kerasapi}
\bibfield{author}{\bibinfo{person}{{Anonymized}}.}
  \bibinfo{year}{2021}\natexlab{}.
\newblock \bibinfo{title}{Keras documentation}.
\newblock
\newblock
\urldef\tempurl%
\url{https://keras.io/api/}
\showURL{%
\tempurl}


\bibitem[\protect\citeauthoryear{Arunava}{Arunava}{2018}]%
        {malaria2018kaggle}
\bibfield{author}{\bibinfo{person}{Arunava}.} \bibinfo{year}{2018}\natexlab{}.
\newblock \bibinfo{title}{Malaria Cell Images Dataset}.
\newblock
\newblock
\urldef\tempurl%
\url{https://www.kaggle.com/iarunava/cell-images-for-detecting-malaria}
\showURL{%
\tempurl}


\bibitem[\protect\citeauthoryear{Baker, Gupta, Naik, and Raskar}{Baker
  et~al\mbox{.}}{2017}]%
        {DBLP:conf/iclr/BakerGNR17}
\bibfield{author}{\bibinfo{person}{Bowen Baker}, \bibinfo{person}{Otkrist
  Gupta}, \bibinfo{person}{Nikhil Naik}, {and} \bibinfo{person}{Ramesh
  Raskar}.} \bibinfo{year}{2017}\natexlab{}.
\newblock \showarticletitle{Designing Neural Network Architectures using
  Reinforcement Learning}. In \bibinfo{booktitle}{\emph{5th International
  Conference on Learning Representations, {ICLR} 2017, Toulon, France, April
  24-26, 2017, Conference Track Proceedings}}.
  \bibinfo{publisher}{OpenReview.net}.
\newblock
\urldef\tempurl%
\url{https://openreview.net/forum?id=S1c2cvqee}
\showURL{%
\tempurl}


\bibitem[\protect\citeauthoryear{Bansal}{Bansal}{2018}]%
        {IIC2018kaggle}
\bibfield{author}{\bibinfo{person}{Puneet Bansal}.}
  \bibinfo{year}{2018}\natexlab{}.
\newblock \bibinfo{booktitle}{\emph{Intel Image Classification}}.
\newblock
\urldef\tempurl%
\url{https://www.kaggle.com/puneet6060/intel-image-classification}
\showURL{%
\tempurl}


\bibitem[\protect\citeauthoryear{Biswas and Rajan}{Biswas and Rajan}{2020}]%
        {biswas20machine}
\bibfield{author}{\bibinfo{person}{Sumon Biswas} {and} \bibinfo{person}{Hridesh
  Rajan}.} \bibinfo{year}{2020}\natexlab{}.
\newblock \showarticletitle{Do the Machine Learning Models on a Crowd Sourced
  Platform Exhibit Bias? An Empirical Study on Model Fairness}. In
  \bibinfo{booktitle}{\emph{ESEC/FSE'2020: The 28th ACM Joint European Software
  Engineering Conference and Symposium on the Foundations of Software
  Engineering}} (Sacramento, California, United States).
\newblock


\bibitem[\protect\citeauthoryear{Biswas, Wardat, and Rajan}{Biswas
  et~al\mbox{.}}{2022}]%
        {biswas22art}
\bibfield{author}{\bibinfo{person}{Sumon Biswas}, \bibinfo{person}{Mohammad
  Wardat}, {and} \bibinfo{person}{Hridesh Rajan}.}
  \bibinfo{year}{2022}\natexlab{}.
\newblock \showarticletitle{The Art and Practice of Data Science Pipelines: A
  Comprehensive Study of Data Science Pipelines In Theory, In-The-Small, and
  In-The-Large}. In \bibinfo{booktitle}{\emph{ICSE'22: The 44th International
  Conference on Software Engineering}} (Pittsburgh, PA, USA).
\newblock


\bibitem[\protect\citeauthoryear{Borges, Hora, and Valente}{Borges
  et~al\mbox{.}}{2016}]%
        {borges2016understanding}
\bibfield{author}{\bibinfo{person}{Hudson Borges}, \bibinfo{person}{Andre
  Hora}, {and} \bibinfo{person}{Marco~Tulio Valente}.}
  \bibinfo{year}{2016}\natexlab{}.
\newblock \showarticletitle{Understanding the factors that impact the
  popularity of GitHub repositories}. In \bibinfo{booktitle}{\emph{2016 IEEE
  International Conference on Software Maintenance and Evolution (ICSME)}}.
  IEEE, \bibinfo{pages}{334--344}.
\newblock


\bibitem[\protect\citeauthoryear{Cai, Chen, Zhang, Yu, and Wang}{Cai
  et~al\mbox{.}}{2018}]%
        {cai2018efficient}
\bibfield{author}{\bibinfo{person}{Han Cai}, \bibinfo{person}{Tianyao Chen},
  \bibinfo{person}{Weinan Zhang}, \bibinfo{person}{Yong Yu}, {and}
  \bibinfo{person}{Jun Wang}.} \bibinfo{year}{2018}\natexlab{}.
\newblock \showarticletitle{Efficient architecture search by network
  transformation}. In \bibinfo{booktitle}{\emph{Thirty-Second AAAI Conference
  on Artificial Intelligence}}.
\newblock


\bibitem[\protect\citeauthoryear{Cambronero, Cito, and Rinard}{Cambronero
  et~al\mbox{.}}{2020}]%
        {cambronero2020ams}
\bibfield{author}{\bibinfo{person}{Jos{\'e}~P Cambronero},
  \bibinfo{person}{J{\"u}rgen Cito}, {and} \bibinfo{person}{Martin~C Rinard}.}
  \bibinfo{year}{2020}\natexlab{}.
\newblock \showarticletitle{Ams: Generating automl search spaces from weak
  specifications}. In \bibinfo{booktitle}{\emph{Proceedings of the 28th ACM
  Joint Meeting on European Software Engineering Conference and Symposium on
  the Foundations of Software Engineering}}. \bibinfo{pages}{763--774}.
\newblock


\bibitem[\protect\citeauthoryear{Cambronero and Rinard}{Cambronero and
  Rinard}{2019}]%
        {cambronero2019autogenerating}
\bibfield{author}{\bibinfo{person}{Jos{\'e}~P Cambronero} {and}
  \bibinfo{person}{Martin~C Rinard}.} \bibinfo{year}{2019}\natexlab{}.
\newblock \showarticletitle{AL: autogenerating supervised learning programs}.
\newblock \bibinfo{journal}{\emph{Proceedings of the ACM on Programming
  Languages}} \bibinfo{volume}{3}, \bibinfo{number}{OOPSLA}
  (\bibinfo{year}{2019}), \bibinfo{pages}{1--28}.
\newblock


\bibitem[\protect\citeauthoryear{Chen, Goodfellow, and Shlens}{Chen
  et~al\mbox{.}}{2016}]%
        {DBLP:journals/corr/ChenGS15}
\bibfield{author}{\bibinfo{person}{Tianqi Chen}, \bibinfo{person}{Ian~J.
  Goodfellow}, {and} \bibinfo{person}{Jonathon Shlens}.}
  \bibinfo{year}{2016}\natexlab{}.
\newblock \showarticletitle{Net2Net: Accelerating Learning via Knowledge
  Transfer}. In \bibinfo{booktitle}{\emph{4th International Conference on
  Learning Representations, {ICLR} 2016, San Juan, Puerto Rico, May 2-4, 2016,
  Conference Track Proceedings}}, \bibfield{editor}{\bibinfo{person}{Yoshua
  Bengio} {and} \bibinfo{person}{Yann LeCun}} (Eds.).
\newblock
\urldef\tempurl%
\url{http://arxiv.org/abs/1511.05641}
\showURL{%
\tempurl}


\bibitem[\protect\citeauthoryear{Chen, Yang, Zhang, MENG, Xiao, and Sun}{Chen
  et~al\mbox{.}}{2019}]%
        {chen2019nips}
\bibfield{author}{\bibinfo{person}{Yukang Chen}, \bibinfo{person}{Tong Yang},
  \bibinfo{person}{Xiangyu Zhang}, \bibinfo{person}{GAOFENG MENG},
  \bibinfo{person}{Xinyu Xiao}, {and} \bibinfo{person}{Jian Sun}.}
  \bibinfo{year}{2019}\natexlab{}.
\newblock \showarticletitle{DetNAS: Backbone Search for Object Detection}. In
  \bibinfo{booktitle}{\emph{Advances in Neural Information Processing
  Systems}}, \bibfield{editor}{\bibinfo{person}{H.~Wallach},
  \bibinfo{person}{H.~Larochelle}, \bibinfo{person}{A.~Beygelzimer},
  \bibinfo{person}{F.~d\textquotesingle Alch\'{e}-Buc},
  \bibinfo{person}{E.~Fox}, {and} \bibinfo{person}{R.~Garnett}} (Eds.),
  Vol.~\bibinfo{volume}{32}. \bibinfo{publisher}{Curran Associates, Inc.}
\newblock
\urldef\tempurl%
\url{https://proceedings.neurips.cc/paper/2019/file/228b25587479f2fc7570428e8bcbabdc-Paper.pdf}
\showURL{%
\tempurl}


\bibitem[\protect\citeauthoryear{Cubuk, Zoph, Mane, Vasudevan, and Le}{Cubuk
  et~al\mbox{.}}{2019}]%
        {cubuk2019autoaugment}
\bibfield{author}{\bibinfo{person}{Ekin~D Cubuk}, \bibinfo{person}{Barret
  Zoph}, \bibinfo{person}{Dandelion Mane}, \bibinfo{person}{Vijay Vasudevan},
  {and} \bibinfo{person}{Quoc~V Le}.} \bibinfo{year}{2019}\natexlab{}.
\newblock \showarticletitle{Autoaugment: Learning augmentation strategies from
  data}. In \bibinfo{booktitle}{\emph{Proceedings of the IEEE/CVF Conference on
  Computer Vision and Pattern Recognition}}. \bibinfo{pages}{113--123}.
\newblock


\bibitem[\protect\citeauthoryear{Cui, Chen, Li, Liu, Shen, and Jia}{Cui
  et~al\mbox{.}}{2019}]%
        {cui2019fast}
\bibfield{author}{\bibinfo{person}{Jiequan Cui}, \bibinfo{person}{Pengguang
  Chen}, \bibinfo{person}{Ruiyu Li}, \bibinfo{person}{Shu Liu},
  \bibinfo{person}{Xiaoyong Shen}, {and} \bibinfo{person}{Jiaya Jia}.}
  \bibinfo{year}{2019}\natexlab{}.
\newblock \showarticletitle{Fast and practical neural architecture search}. In
  \bibinfo{booktitle}{\emph{Proceedings of the IEEE/CVF International
  Conference on Computer Vision}}. \bibinfo{pages}{6509--6518}.
\newblock


\bibitem[\protect\citeauthoryear{Dilhara, Ketkar, Sannidhi, and Dig}{Dilhara
  et~al\mbox{.}}{2022}]%
        {DilharaICSE2022RepetitiveCode}
\bibfield{author}{\bibinfo{person}{Malinda Dilhara}, \bibinfo{person}{Ameya
  Ketkar}, \bibinfo{person}{Nikhith Sannidhi}, {and} \bibinfo{person}{Danny
  Dig}.} \bibinfo{year}{2022}\natexlab{}.
\newblock \showarticletitle{Discovering Repetitive Code Changes in Python ML
  Systems}. In \bibinfo{booktitle}{\emph{International Conference on Software
  Engineering}} (Pittsburgh, United States) \emph{(\bibinfo{series}{ICSE
  '22})}. ACM/IEEE.
\newblock
\urldef\tempurl%
\url{https://doi.org/10.1145/3510003.3510225}
\showDOI{\tempurl}
\newblock
\shownote{To appear.}


\bibitem[\protect\citeauthoryear{Elsken, Metzen, and Hutter}{Elsken
  et~al\mbox{.}}{2018}]%
        {elsken2018simple}
\bibfield{author}{\bibinfo{person}{Thomas Elsken}, \bibinfo{person}{Jan~Hendrik
  Metzen}, {and} \bibinfo{person}{Frank Hutter}.}
  \bibinfo{year}{2018}\natexlab{}.
\newblock \bibinfo{title}{Simple and efficient architecture search for
  Convolutional Neural Networks}.
\newblock
\newblock
\urldef\tempurl%
\url{https://openreview.net/forum?id=SySaJ0xCZ}
\showURL{%
\tempurl}


\bibitem[\protect\citeauthoryear{Elsken, Metzen, and Hutter}{Elsken
  et~al\mbox{.}}{2019}]%
        {elsken2018efficient}
\bibfield{author}{\bibinfo{person}{Thomas Elsken}, \bibinfo{person}{Jan~Hendrik
  Metzen}, {and} \bibinfo{person}{Frank Hutter}.}
  \bibinfo{year}{2019}\natexlab{}.
\newblock \showarticletitle{Efficient Multi-Objective Neural Architecture
  Search via Lamarckian Evolution}. In \bibinfo{booktitle}{\emph{International
  Conference on Learning Representations}}.
\newblock
\urldef\tempurl%
\url{https://openreview.net/forum?id=ByME42AqK7}
\showURL{%
\tempurl}


\bibitem[\protect\citeauthoryear{Feurer, Eggensperger, Falkner, Lindauer, and
  Hutter}{Feurer et~al\mbox{.}}{2020}]%
        {feurer2020auto}
\bibfield{author}{\bibinfo{person}{Matthias Feurer}, \bibinfo{person}{Katharina
  Eggensperger}, \bibinfo{person}{Stefan Falkner}, \bibinfo{person}{Marius
  Lindauer}, {and} \bibinfo{person}{Frank Hutter}.}
  \bibinfo{year}{2020}\natexlab{}.
\newblock \showarticletitle{Auto-sklearn 2.0: The next generation}.
\newblock \bibinfo{journal}{\emph{arXiv preprint arXiv:2007.04074}}
  (\bibinfo{year}{2020}).
\newblock


\bibitem[\protect\citeauthoryear{Feurer, Klein, Eggensperger, Springenberg,
  Blum, and Hutter}{Feurer et~al\mbox{.}}{2015}]%
        {feurer2015efficient}
\bibfield{author}{\bibinfo{person}{Matthias Feurer}, \bibinfo{person}{Aaron
  Klein}, \bibinfo{person}{Katharina Eggensperger}, \bibinfo{person}{Jost
  Springenberg}, \bibinfo{person}{Manuel Blum}, {and} \bibinfo{person}{Frank
  Hutter}.} \bibinfo{year}{2015}\natexlab{}.
\newblock \showarticletitle{Efficient and robust automated machine learning}.
  In \bibinfo{booktitle}{\emph{Advances in neural information processing
  systems}}. \bibinfo{pages}{2962--2970}.
\newblock


\bibitem[\protect\citeauthoryear{{Giang Nguyen, Md Johirul Islam, Rangeet Pan,
  and Hridesh Rajan}}{{Giang Nguyen, Md Johirul Islam, Rangeet Pan, and Hridesh
  Rajan}}{2021}]%
        {mnartifact}
\bibfield{author}{\bibinfo{person}{{Giang Nguyen, Md Johirul Islam, Rangeet
  Pan, and Hridesh Rajan}}.} \bibinfo{year}{2021}\natexlab{}.
\newblock \bibinfo{title}{Manas artifact}.
\newblock
\newblock
\urldef\tempurl%
\url{https://github.com/giangnm58/Manas}
\showURL{%
\tempurl}


\bibitem[\protect\citeauthoryear{Hall, Frank, Holmes, Pfahringer, Reutemann,
  and Witten}{Hall et~al\mbox{.}}{2009}]%
        {hall2009weka}
\bibfield{author}{\bibinfo{person}{Mark Hall}, \bibinfo{person}{Eibe Frank},
  \bibinfo{person}{Geoffrey Holmes}, \bibinfo{person}{Bernhard Pfahringer},
  \bibinfo{person}{Peter Reutemann}, {and} \bibinfo{person}{Ian~H Witten}.}
  \bibinfo{year}{2009}\natexlab{}.
\newblock \showarticletitle{The WEKA data mining software: an update}.
\newblock \bibinfo{journal}{\emph{ACM SIGKDD explorations newsletter}}
  \bibinfo{volume}{11}, \bibinfo{number}{1} (\bibinfo{year}{2009}),
  \bibinfo{pages}{10--18}.
\newblock


\bibitem[\protect\citeauthoryear{Hamerly and Elkan}{Hamerly and Elkan}{2004}]%
        {hamerly2004learning}
\bibfield{author}{\bibinfo{person}{Greg Hamerly} {and} \bibinfo{person}{Charles
  Elkan}.} \bibinfo{year}{2004}\natexlab{}.
\newblock \showarticletitle{Learning the k in k-means}. In
  \bibinfo{booktitle}{\emph{Advances in neural information processing
  systems}}. \bibinfo{pages}{281--288}.
\newblock


\bibitem[\protect\citeauthoryear{Han, Mao, and Dally}{Han
  et~al\mbox{.}}{2016}]%
        {han2015deep}
\bibfield{author}{\bibinfo{person}{Song Han}, \bibinfo{person}{Huizi Mao},
  {and} \bibinfo{person}{William~J. Dally}.} \bibinfo{year}{2016}\natexlab{}.
\newblock \showarticletitle{Deep Compression: Compressing Deep Neural Network
  with Pruning, Trained Quantization and Huffman Coding}. In
  \bibinfo{booktitle}{\emph{4th International Conference on Learning
  Representations, {ICLR} 2016, San Juan, Puerto Rico, May 2-4, 2016,
  Conference Track Proceedings}}, \bibfield{editor}{\bibinfo{person}{Yoshua
  Bengio} {and} \bibinfo{person}{Yann LeCun}} (Eds.).
\newblock
\urldef\tempurl%
\url{http://arxiv.org/abs/1510.00149}
\showURL{%
\tempurl}


\bibitem[\protect\citeauthoryear{He, Zhang, Ren, and Sun}{He
  et~al\mbox{.}}{2016}]%
        {he2016deep}
\bibfield{author}{\bibinfo{person}{Kaiming He}, \bibinfo{person}{Xiangyu
  Zhang}, \bibinfo{person}{Shaoqing Ren}, {and} \bibinfo{person}{Jian Sun}.}
  \bibinfo{year}{2016}\natexlab{}.
\newblock \showarticletitle{Deep residual learning for image recognition}. In
  \bibinfo{booktitle}{\emph{Proceedings of the IEEE conference on computer
  vision and pattern recognition}}. \bibinfo{pages}{770--778}.
\newblock


\bibitem[\protect\citeauthoryear{Huang, Liu, Van Der~Maaten, and
  Weinberger}{Huang et~al\mbox{.}}{2017}]%
        {huang2017densely}
\bibfield{author}{\bibinfo{person}{Gao Huang}, \bibinfo{person}{Zhuang Liu},
  \bibinfo{person}{Laurens Van Der~Maaten}, {and} \bibinfo{person}{Kilian~Q
  Weinberger}.} \bibinfo{year}{2017}\natexlab{}.
\newblock \showarticletitle{Densely connected convolutional networks}. In
  \bibinfo{booktitle}{\emph{Proceedings of the IEEE conference on computer
  vision and pattern recognition}}. \bibinfo{pages}{4700--4708}.
\newblock


\bibitem[\protect\citeauthoryear{Iandola, Moskewicz, Ashraf, and
  Keutzer}{Iandola et~al\mbox{.}}{2016}]%
        {iandola2016firecaffe}
\bibfield{author}{\bibinfo{person}{Forrest~N Iandola},
  \bibinfo{person}{Matthew~W Moskewicz}, \bibinfo{person}{Khalid Ashraf}, {and}
  \bibinfo{person}{Kurt Keutzer}.} \bibinfo{year}{2016}\natexlab{}.
\newblock \showarticletitle{Firecaffe: near-linear acceleration of deep neural
  network training on compute clusters}. In
  \bibinfo{booktitle}{\emph{Proceedings of the IEEE Conference on Computer
  Vision and Pattern Recognition}}. \bibinfo{pages}{2592--2600}.
\newblock


\bibitem[\protect\citeauthoryear{Ioffe and Szegedy}{Ioffe and Szegedy}{2015}]%
        {ioffe2015batch}
\bibfield{author}{\bibinfo{person}{Sergey Ioffe} {and}
  \bibinfo{person}{Christian Szegedy}.} \bibinfo{year}{2015}\natexlab{}.
\newblock \showarticletitle{Batch normalization: Accelerating deep network
  training by reducing internal covariate shift}. In
  \bibinfo{booktitle}{\emph{International conference on machine learning}}.
  PMLR, \bibinfo{pages}{448--456}.
\newblock


\bibitem[\protect\citeauthoryear{Islam, Nguyen, Pan, and Rajan}{Islam
  et~al\mbox{.}}{2019}]%
        {islam19}
\bibfield{author}{\bibinfo{person}{Md~Johirul Islam}, \bibinfo{person}{Giang
  Nguyen}, \bibinfo{person}{Rangeet Pan}, {and} \bibinfo{person}{Hridesh
  Rajan}.} \bibinfo{year}{2019}\natexlab{}.
\newblock \showarticletitle{A Comprehensive Study on Deep Learning Bug
  Characteristics}. In \bibinfo{booktitle}{\emph{ESEC/FSE'19: The ACM Joint
  European Software Engineering Conference and Symposium on the Foundations of
  Software Engineering (ESEC/FSE)}} \emph{(\bibinfo{series}{ESEC/FSE 2019})}.
\newblock


\bibitem[\protect\citeauthoryear{Islam, Pan, Nguyen, and Rajan}{Islam
  et~al\mbox{.}}{2020}]%
        {islam20repairing}
\bibfield{author}{\bibinfo{person}{Md~Johirul Islam}, \bibinfo{person}{Rangeet
  Pan}, \bibinfo{person}{Giang Nguyen}, {and} \bibinfo{person}{Hridesh Rajan}.}
  \bibinfo{year}{2020}\natexlab{}.
\newblock \showarticletitle{Repairing Deep Neural Networks: Fix Patterns and
  Challenges}. In \bibinfo{booktitle}{\emph{ICSE'20: The 42nd International
  Conference on Software Engineering}} (Seoul, South Korea).
\newblock


\bibitem[\protect\citeauthoryear{Jin, Song, and Hu}{Jin et~al\mbox{.}}{2019}]%
        {jin2019auto}
\bibfield{author}{\bibinfo{person}{Haifeng Jin}, \bibinfo{person}{Qingquan
  Song}, {and} \bibinfo{person}{Xia Hu}.} \bibinfo{year}{2019}\natexlab{}.
\newblock \showarticletitle{Auto-Keras: An Efficient Neural Architecture Search
  System}. In \bibinfo{booktitle}{\emph{Proceedings of the 25th ACM SIGKDD
  International Conference on Knowledge Discovery \& Data Mining}}. ACM,
  \bibinfo{pages}{1946--1956}.
\newblock


\bibitem[\protect\citeauthoryear{Kandasamy, Neiswanger, Schneider, Poczos, and
  Xing}{Kandasamy et~al\mbox{.}}{2018}]%
        {kandasamy2018neural}
\bibfield{author}{\bibinfo{person}{Kirthevasan Kandasamy},
  \bibinfo{person}{Willie Neiswanger}, \bibinfo{person}{Jeff Schneider},
  \bibinfo{person}{Barnabas Poczos}, {and} \bibinfo{person}{Eric~P Xing}.}
  \bibinfo{year}{2018}\natexlab{}.
\newblock \showarticletitle{Neural Architecture Search with Bayesian
  Optimisation and Optimal Transport}. In \bibinfo{booktitle}{\emph{Advances in
  Neural Information Processing Systems}},
  \bibfield{editor}{\bibinfo{person}{S.~Bengio}, \bibinfo{person}{H.~Wallach},
  \bibinfo{person}{H.~Larochelle}, \bibinfo{person}{K.~Grauman},
  \bibinfo{person}{N.~Cesa-Bianchi}, {and} \bibinfo{person}{R.~Garnett}}
  (Eds.), Vol.~\bibinfo{volume}{31}. \bibinfo{publisher}{Curran Associates,
  Inc.}
\newblock
\urldef\tempurl%
\url{https://proceedings.neurips.cc/paper/2018/file/f33ba15effa5c10e873bf3842afb46a6-Paper.pdf}
\showURL{%
\tempurl}


\bibitem[\protect\citeauthoryear{Krizhevsky, Hinton, et~al\mbox{.}}{Krizhevsky
  et~al\mbox{.}}{2009}]%
        {krizhevsky2009learning}
\bibfield{author}{\bibinfo{person}{Alex Krizhevsky}, \bibinfo{person}{Geoffrey
  Hinton}, {et~al\mbox{.}}} \bibinfo{year}{2009}\natexlab{}.
\newblock \bibinfo{booktitle}{\emph{Learning multiple layers of features from
  tiny images}}.
\newblock \bibinfo{type}{{T}echnical {R}eport}.
  \bibinfo{institution}{Citeseer}.
\newblock


\bibitem[\protect\citeauthoryear{LeCun, Bottou, Bengio, Haffner,
  et~al\mbox{.}}{LeCun et~al\mbox{.}}{1998}]%
        {lecun1998gradient}
\bibfield{author}{\bibinfo{person}{Yann LeCun}, \bibinfo{person}{L{\'e}on
  Bottou}, \bibinfo{person}{Yoshua Bengio}, \bibinfo{person}{Patrick Haffner},
  {et~al\mbox{.}}} \bibinfo{year}{1998}\natexlab{}.
\newblock \showarticletitle{Gradient-based learning applied to document
  recognition}.
\newblock \bibinfo{journal}{\emph{Proc. IEEE}} \bibinfo{volume}{86},
  \bibinfo{number}{11} (\bibinfo{year}{1998}), \bibinfo{pages}{2278--2324}.
\newblock


\bibitem[\protect\citeauthoryear{Lin, Chen, and Yan}{Lin et~al\mbox{.}}{2014}]%
        {lin2013network}
\bibfield{author}{\bibinfo{person}{Min Lin}, \bibinfo{person}{Qiang Chen},
  {and} \bibinfo{person}{Shuicheng Yan}.} \bibinfo{year}{2014}\natexlab{}.
\newblock \showarticletitle{Network In Network}. In
  \bibinfo{booktitle}{\emph{2nd International Conference on Learning
  Representations, {ICLR} 2014, Banff, AB, Canada, April 14-16, 2014,
  Conference Track Proceedings}}, \bibfield{editor}{\bibinfo{person}{Yoshua
  Bengio} {and} \bibinfo{person}{Yann LeCun}} (Eds.).
\newblock
\urldef\tempurl%
\url{http://arxiv.org/abs/1312.4400}
\showURL{%
\tempurl}


\bibitem[\protect\citeauthoryear{Mader}{Mader}{2018}]%
        {ham2018kaggle}
\bibfield{author}{\bibinfo{person}{Kevin Mader}.}
  \bibinfo{year}{2018}\natexlab{}.
\newblock \bibinfo{title}{Skin Cancer MNIST: HAM10000}.
\newblock
\newblock
\urldef\tempurl%
\url{https://www.kaggle.com/kmader/skin-cancer-mnist-ham10000}
\showURL{%
\tempurl}


\bibitem[\protect\citeauthoryear{Mamaev}{Mamaev}{2017}]%
        {flower2017kaggle}
\bibfield{author}{\bibinfo{person}{Alexander Mamaev}.}
  \bibinfo{year}{2017}\natexlab{}.
\newblock \bibinfo{title}{Flowers Recognition}.
\newblock
\newblock
\urldef\tempurl%
\url{https://www.kaggle.com/alxmamaev/flowers-recognition}
\showURL{%
\tempurl}


\bibitem[\protect\citeauthoryear{Mavi}{Mavi}{2017}]%
        {sd2017kaggle}
\bibfield{author}{\bibinfo{person}{Arda Mavi}.}
  \bibinfo{year}{2017}\natexlab{}.
\newblock \bibinfo{title}{Sign Language Digits Dataset}.
\newblock
\newblock
\urldef\tempurl%
\url{https://www.kaggle.com/ardamavi/sign-language-digits-dataset}
\showURL{%
\tempurl}


\bibitem[\protect\citeauthoryear{Mooney}{Mooney}{2017a}]%
        {bloodcell2017kaggle}
\bibfield{author}{\bibinfo{person}{Paul Mooney}.}
  \bibinfo{year}{2017}\natexlab{a}.
\newblock \bibinfo{title}{Blood Cell Images}.
\newblock
\newblock
\urldef\tempurl%
\url{https://www.kaggle.com/paultimothymooney/blood-cells}
\showURL{%
\tempurl}


\bibitem[\protect\citeauthoryear{Mooney}{Mooney}{2017b}]%
        {breast2017kaggle}
\bibfield{author}{\bibinfo{person}{Paul Mooney}.}
  \bibinfo{year}{2017}\natexlab{b}.
\newblock \bibinfo{booktitle}{\emph{Breast Histopathology Images}}.
\newblock
\urldef\tempurl%
\url{https://www.kaggle.com/paultimothymooney/breast-histopathology-images}
\showURL{%
\tempurl}


\bibitem[\protect\citeauthoryear{Nayman, Noy, Ridnik, Friedman, Jin, and
  Zelnik}{Nayman et~al\mbox{.}}{2019}]%
        {nayman2019nips}
\bibfield{author}{\bibinfo{person}{Niv Nayman}, \bibinfo{person}{Asaf Noy},
  \bibinfo{person}{Tal Ridnik}, \bibinfo{person}{Itamar Friedman},
  \bibinfo{person}{Rong Jin}, {and} \bibinfo{person}{Lihi Zelnik}.}
  \bibinfo{year}{2019}\natexlab{}.
\newblock \showarticletitle{XNAS: Neural Architecture Search with Expert
  Advice}. In \bibinfo{booktitle}{\emph{Advances in Neural Information
  Processing Systems}}, \bibfield{editor}{\bibinfo{person}{H.~Wallach},
  \bibinfo{person}{H.~Larochelle}, \bibinfo{person}{A.~Beygelzimer},
  \bibinfo{person}{F.~d\textquotesingle Alch\'{e}-Buc},
  \bibinfo{person}{E.~Fox}, {and} \bibinfo{person}{R.~Garnett}} (Eds.),
  Vol.~\bibinfo{volume}{32}. \bibinfo{publisher}{Curran Associates, Inc.}
\newblock
\urldef\tempurl%
\url{https://proceedings.neurips.cc/paper/2019/file/00e26af6ac3b1c1c49d7c3d79c60d000-Paper.pdf}
\showURL{%
\tempurl}


\bibitem[\protect\citeauthoryear{Olson, Bartley, Urbanowicz, and Moore}{Olson
  et~al\mbox{.}}{2016a}]%
        {olson2016evaluation}
\bibfield{author}{\bibinfo{person}{Randal~S Olson}, \bibinfo{person}{Nathan
  Bartley}, \bibinfo{person}{Ryan~J Urbanowicz}, {and} \bibinfo{person}{Jason~H
  Moore}.} \bibinfo{year}{2016}\natexlab{a}.
\newblock \showarticletitle{Evaluation of a tree-based pipeline optimization
  tool for automating data science}. In \bibinfo{booktitle}{\emph{Proceedings
  of the Genetic and Evolutionary Computation Conference 2016}}. ACM,
  \bibinfo{pages}{485--492}.
\newblock


\bibitem[\protect\citeauthoryear{Olson, Urbanowicz, Andrews, Lavender, Moore,
  et~al\mbox{.}}{Olson et~al\mbox{.}}{2016b}]%
        {olson2016automating}
\bibfield{author}{\bibinfo{person}{Randal~S Olson}, \bibinfo{person}{Ryan~J
  Urbanowicz}, \bibinfo{person}{Peter~C Andrews}, \bibinfo{person}{Nicole~A
  Lavender}, \bibinfo{person}{Jason~H Moore}, {et~al\mbox{.}}}
  \bibinfo{year}{2016}\natexlab{b}.
\newblock \showarticletitle{Automating biomedical data science through
  tree-based pipeline optimization}. In \bibinfo{booktitle}{\emph{European
  Conference on the Applications of Evolutionary Computation}}. Springer,
  \bibinfo{pages}{123--137}.
\newblock


\bibitem[\protect\citeauthoryear{Pan and Rajan}{Pan and Rajan}{2022}]%
        {pan22decomposing}
\bibfield{author}{\bibinfo{person}{Rangeet Pan} {and} \bibinfo{person}{Hridesh
  Rajan}.} \bibinfo{year}{2022}\natexlab{}.
\newblock \showarticletitle{Decomposing Convolutional Neural Networks into
  Reusable and Replaceable Modules}. In \bibinfo{booktitle}{\emph{ICSE'22: The
  44th International Conference on Software Engineering}} (Pittsburgh, PA,
  USA).
\newblock


\bibitem[\protect\citeauthoryear{Peng, Sun, ZHANG, Tan, and Yan}{Peng
  et~al\mbox{.}}{2019}]%
        {peng2019nips}
\bibfield{author}{\bibinfo{person}{Junran Peng}, \bibinfo{person}{Ming Sun},
  \bibinfo{person}{ZHAO-XIANG ZHANG}, \bibinfo{person}{Tieniu Tan}, {and}
  \bibinfo{person}{Junjie Yan}.} \bibinfo{year}{2019}\natexlab{}.
\newblock \showarticletitle{Efficient Neural Architecture Transformation Search
  in Channel-Level for Object Detection}. In \bibinfo{booktitle}{\emph{Advances
  in Neural Information Processing Systems}},
  \bibfield{editor}{\bibinfo{person}{H.~Wallach},
  \bibinfo{person}{H.~Larochelle}, \bibinfo{person}{A.~Beygelzimer},
  \bibinfo{person}{F.~d\textquotesingle Alch\'{e}-Buc},
  \bibinfo{person}{E.~Fox}, {and} \bibinfo{person}{R.~Garnett}} (Eds.),
  Vol.~\bibinfo{volume}{32}. \bibinfo{publisher}{Curran Associates, Inc.}
\newblock
\urldef\tempurl%
\url{https://proceedings.neurips.cc/paper/2019/file/3aaa3db6a8983226601cac5dde15a26b-Paper.pdf}
\showURL{%
\tempurl}


\bibitem[\protect\citeauthoryear{Real, Moore, Selle, Saxena, Suematsu, Tan, Le,
  and Kurakin}{Real et~al\mbox{.}}{2017}]%
        {real2017large}
\bibfield{author}{\bibinfo{person}{Esteban Real}, \bibinfo{person}{Sherry
  Moore}, \bibinfo{person}{Andrew Selle}, \bibinfo{person}{Saurabh Saxena},
  \bibinfo{person}{Yutaka~Leon Suematsu}, \bibinfo{person}{Jie Tan},
  \bibinfo{person}{Quoc~V Le}, {and} \bibinfo{person}{Alexey Kurakin}.}
  \bibinfo{year}{2017}\natexlab{}.
\newblock \showarticletitle{Large-scale evolution of image classifiers}. In
  \bibinfo{booktitle}{\emph{Proceedings of the 34th International Conference on
  Machine Learning-Volume 70}}. JMLR. org, \bibinfo{pages}{2902--2911}.
\newblock


\bibitem[\protect\citeauthoryear{Rosebrock}{Rosebrock}{2019}]%
        {rosebrock2019ak}
\bibfield{author}{\bibinfo{person}{Adrian Rosebrock}.}
  \bibinfo{year}{2019}\natexlab{}.
\newblock \bibinfo{title}{Auto-Keras and AutoML: A Getting Started Guide}.
\newblock
\newblock
\urldef\tempurl%
\url{https://www.pyimagesearch.com/2019/01/07/auto-keras-and-automl-a-getting-started-guide}
\showURL{%
\tempurl}


\bibitem[\protect\citeauthoryear{Srivastava, Hinton, Krizhevsky, Sutskever, and
  Salakhutdinov}{Srivastava et~al\mbox{.}}{2014}]%
        {srivastava2014dropout}
\bibfield{author}{\bibinfo{person}{Nitish Srivastava},
  \bibinfo{person}{Geoffrey Hinton}, \bibinfo{person}{Alex Krizhevsky},
  \bibinfo{person}{Ilya Sutskever}, {and} \bibinfo{person}{Ruslan
  Salakhutdinov}.} \bibinfo{year}{2014}\natexlab{}.
\newblock \showarticletitle{Dropout: a simple way to prevent neural networks
  from overfitting}.
\newblock \bibinfo{journal}{\emph{The journal of machine learning research}}
  \bibinfo{volume}{15}, \bibinfo{number}{1} (\bibinfo{year}{2014}),
  \bibinfo{pages}{1929--1958}.
\newblock


\bibitem[\protect\citeauthoryear{Suganuma, Shirakawa, and Nagao}{Suganuma
  et~al\mbox{.}}{2017}]%
        {suganuma2017genetic}
\bibfield{author}{\bibinfo{person}{Masanori Suganuma},
  \bibinfo{person}{Shinichi Shirakawa}, {and} \bibinfo{person}{Tomoharu
  Nagao}.} \bibinfo{year}{2017}\natexlab{}.
\newblock \showarticletitle{A genetic programming approach to designing
  convolutional neural network architectures}. In
  \bibinfo{booktitle}{\emph{Proceedings of the Genetic and Evolutionary
  Computation Conference}}. ACM, \bibinfo{pages}{497--504}.
\newblock


\bibitem[\protect\citeauthoryear{Tan and Le}{Tan and Le}{2019}]%
        {tan2019efficientnet}
\bibfield{author}{\bibinfo{person}{Mingxing Tan} {and} \bibinfo{person}{Quoc
  Le}.} \bibinfo{year}{2019}\natexlab{}.
\newblock \showarticletitle{Efficientnet: Rethinking model scaling for
  convolutional neural networks}. In \bibinfo{booktitle}{\emph{International
  Conference on Machine Learning}}. PMLR, \bibinfo{pages}{6105--6114}.
\newblock


\bibitem[\protect\citeauthoryear{Tecperson}{Tecperson}{2017}]%
        {sl2017kaggle}
\bibfield{author}{\bibinfo{person}{Tecperson}.}
  \bibinfo{year}{2017}\natexlab{}.
\newblock \bibinfo{booktitle}{\emph{Sign Language MNIST}}.
\newblock
\urldef\tempurl%
\url{https://www.kaggle.com/datamunge/sign-language-mnist}
\showURL{%
\tempurl}


\bibitem[\protect\citeauthoryear{Thornton, Hutter, Hoos, and
  Leyton-Brown}{Thornton et~al\mbox{.}}{2013}]%
        {thornton2013auto}
\bibfield{author}{\bibinfo{person}{Chris Thornton}, \bibinfo{person}{Frank
  Hutter}, \bibinfo{person}{Holger~H Hoos}, {and} \bibinfo{person}{Kevin
  Leyton-Brown}.} \bibinfo{year}{2013}\natexlab{}.
\newblock \showarticletitle{Auto-WEKA: Combined selection and hyperparameter
  optimization of classification algorithms}. In
  \bibinfo{booktitle}{\emph{Proceedings of the 19th ACM SIGKDD international
  conference on Knowledge discovery and data mining}}. ACM,
  \bibinfo{pages}{847--855}.
\newblock


\bibitem[\protect\citeauthoryear{Wardat, Cruz, Le, and Rajan}{Wardat
  et~al\mbox{.}}{2022}]%
        {wardat2021deepdiagnosis}
\bibfield{author}{\bibinfo{person}{Mohammad Wardat},
  \bibinfo{person}{Breno~Dantas Cruz}, \bibinfo{person}{Wei Le}, {and}
  \bibinfo{person}{Hridesh Rajan}.} \bibinfo{year}{2022}\natexlab{}.
\newblock \showarticletitle{DeepDiagnosis: Automatically Diagnosing Faults and
  Recommending Actionable Fixes in Deep Learning Programs}. In
  \bibinfo{booktitle}{\emph{ICSE'22: The 44th International Conference on
  Software Engineering}} (Pittsburgh, PA, USA).
\newblock


\bibitem[\protect\citeauthoryear{Wardat, Le, and Rajan}{Wardat
  et~al\mbox{.}}{2021}]%
        {wardat21deeplocalize}
\bibfield{author}{\bibinfo{person}{Mohammad Wardat}, \bibinfo{person}{Wei Le},
  {and} \bibinfo{person}{Hridesh Rajan}.} \bibinfo{year}{2021}\natexlab{}.
\newblock \showarticletitle{DeepLocalize: Fault Localization for Deep Neural
  Networks}. In \bibinfo{booktitle}{\emph{ICSE'21: The 43nd International
  Conference on Software Engineering}} (Virtual Conference).
\newblock


\bibitem[\protect\citeauthoryear{Wei, Wang, Rui, and Chen}{Wei
  et~al\mbox{.}}{2016}]%
        {wei2016network}
\bibfield{author}{\bibinfo{person}{Tao Wei}, \bibinfo{person}{Changhu Wang},
  \bibinfo{person}{Yong Rui}, {and} \bibinfo{person}{Chang~Wen Chen}.}
  \bibinfo{year}{2016}\natexlab{}.
\newblock \showarticletitle{Network morphism}. In
  \bibinfo{booktitle}{\emph{International Conference on Machine Learning}}.
  \bibinfo{pages}{564--572}.
\newblock


\bibitem[\protect\citeauthoryear{Witten, Frank, Hall, and Pal}{Witten
  et~al\mbox{.}}{2016}]%
        {witten2016data}
\bibfield{author}{\bibinfo{person}{Ian~H Witten}, \bibinfo{person}{Eibe Frank},
  \bibinfo{person}{Mark~A Hall}, {and} \bibinfo{person}{Christopher~J Pal}.}
  \bibinfo{year}{2016}\natexlab{}.
\newblock \bibinfo{booktitle}{\emph{Data Mining: Practical machine learning
  tools and techniques}}.
\newblock \bibinfo{publisher}{Morgan Kaufmann}.
\newblock


\bibitem[\protect\citeauthoryear{Xiao, Rasul, and Vollgraf}{Xiao
  et~al\mbox{.}}{2017}]%
        {xiao2017fashion}
\bibfield{author}{\bibinfo{person}{Han Xiao}, \bibinfo{person}{Kashif Rasul},
  {and} \bibinfo{person}{Roland Vollgraf}.} \bibinfo{year}{2017}\natexlab{}.
\newblock \showarticletitle{Fashion-mnist: a novel image dataset for
  benchmarking machine learning algorithms}.
\newblock \bibinfo{journal}{\emph{arXiv preprint arXiv:1708.07747}}
  (\bibinfo{year}{2017}).
\newblock


\bibitem[\protect\citeauthoryear{Xie and Yuille}{Xie and Yuille}{2017}]%
        {xie2017genetic}
\bibfield{author}{\bibinfo{person}{Lingxi Xie} {and} \bibinfo{person}{Alan
  Yuille}.} \bibinfo{year}{2017}\natexlab{}.
\newblock \showarticletitle{Genetic cnn}. In
  \bibinfo{booktitle}{\emph{Proceedings of the IEEE International Conference on
  Computer Vision}}. \bibinfo{pages}{1379--1388}.
\newblock


\bibitem[\protect\citeauthoryear{Zhong, Yan, Wu, Shao, and Liu}{Zhong
  et~al\mbox{.}}{2018}]%
        {zhong2018practical}
\bibfield{author}{\bibinfo{person}{Zhao Zhong}, \bibinfo{person}{Junjie Yan},
  \bibinfo{person}{Wei Wu}, \bibinfo{person}{Jing Shao}, {and}
  \bibinfo{person}{Cheng-Lin Liu}.} \bibinfo{year}{2018}\natexlab{}.
\newblock \showarticletitle{Practical block-wise neural network architecture
  generation}. In \bibinfo{booktitle}{\emph{Proceedings of the IEEE Conference
  on Computer Vision and Pattern Recognition}}. \bibinfo{pages}{2423--2432}.
\newblock


\bibitem[\protect\citeauthoryear{Zhu, Tu, and {Xiangji Huang}}{Zhu
  et~al\mbox{.}}{2020}]%
        {ZHU2020125}
\bibfield{author}{\bibinfo{person}{Runjie Zhu}, \bibinfo{person}{Xinhui Tu},
  {and} \bibinfo{person}{Jimmy {Xiangji Huang}}.}
  \bibinfo{year}{2020}\natexlab{}.
\newblock \showarticletitle{Chapter seven - Deep learning on information
  retrieval and its applications}.
\newblock In \bibinfo{booktitle}{\emph{Deep Learning for Data Analytics}},
  \bibfield{editor}{\bibinfo{person}{Himansu Das},
  \bibinfo{person}{Chittaranjan Pradhan}, {and} \bibinfo{person}{Nilanjan Dey}}
  (Eds.). \bibinfo{publisher}{Academic Press}, \bibinfo{pages}{125--153}.
\newblock
\showISBNx{978-0-12-819764-6}
\urldef\tempurl%
\url{https://doi.org/10.1016/B978-0-12-819764-6.00008-9}
\showDOI{\tempurl}


\bibitem[\protect\citeauthoryear{Zoph and Le}{Zoph and Le}{2017}]%
        {DBLP:conf/iclr/ZophL17}
\bibfield{author}{\bibinfo{person}{Barret Zoph} {and} \bibinfo{person}{Quoc~V.
  Le}.} \bibinfo{year}{2017}\natexlab{}.
\newblock \showarticletitle{Neural Architecture Search with Reinforcement
  Learning}. In \bibinfo{booktitle}{\emph{5th International Conference on
  Learning Representations, {ICLR} 2017, Toulon, France, April 24-26, 2017,
  Conference Track Proceedings}}. \bibinfo{publisher}{OpenReview.net}.
\newblock
\urldef\tempurl%
\url{https://openreview.net/forum?id=r1Ue8Hcxg}
\showURL{%
\tempurl}


\bibitem[\protect\citeauthoryear{Zoph, Vasudevan, Shlens, and Le}{Zoph
  et~al\mbox{.}}{2018}]%
        {zoph2018learning}
\bibfield{author}{\bibinfo{person}{Barret Zoph}, \bibinfo{person}{Vijay
  Vasudevan}, \bibinfo{person}{Jonathon Shlens}, {and} \bibinfo{person}{Quoc~V
  Le}.} \bibinfo{year}{2018}\natexlab{}.
\newblock \showarticletitle{Learning transferable architectures for scalable
  image recognition}. In \bibinfo{booktitle}{\emph{Proceedings of the IEEE
  conference on computer vision and pattern recognition}}.
  \bibinfo{pages}{8697--8710}.
\newblock


\end{thebibliography}
	
	
\end{document}